\documentclass[eqsecnum,preprint,prd,aps,nofootinbib]{revtex4}
\usepackage{epsfig}
\usepackage{amssymb}
\usepackage[dvips]{color}
\usepackage[latin1]{inputenc}
\usepackage{graphicx}

\newcommand{\nc}{\newcommand}
\nc{\be}{\begin{equation}} \nc{\ee}{\end{equation}}
\nc{\bea}{\begin{eqnarray}} \nc{\eea}{\end{eqnarray}}
\nc{\bt}{\begin{tabular}} \nc{\et}{\end{tabular}}
\nc{\ba}{\begin{array}} \nc{\ea}{\end{array}}
\nc{\dy}{\displaystyle} \nc{\pr}{{\rm I}} \nc{\se}{{\rm II}}
\nc{\w}{{\rm w}} \nc{\s}{{\rm s}} \nc{\um}{{1\over 2}}
\nc{\Hc}{{{\cal H}_{\rm c}}}
\def\a{\alpha}

\def\ep{\epsilon}
\def\d{\delta}
\def\g{\gamma}
\def\k{\kappa}
\def\l{\lambda}
\def\s{\sigma}

\def\om{\omega}
\def\p{\partial}
\def\th{\theta}

\def\L{\Lambda}

\def\hA{\hat{A}}
\def\hD{\hat{D}}

\def\hL{\hat{\Lambda}}

\def\hd{\hat{\delta}}
\def\del{\partial}

\def\we{\wedge}

\def\westar{\we_\star}

\def\omtilde{\tilde \om}

\def\rtilde{\tilde r}

\begin{document}
\begin{center}
\bibliographystyle{article}
{\Large \textsc{The Seiberg--Witten map for
non-commutative pure gravity and vacuum Maxwell theory}}
\end{center}

\date{\today}

\author{Elisabetta Di Grezia$^{1}$
\thanks{Electronic address: digrezia@na.infn.it},
Giampiero Esposito$^{1}$
\thanks{Electronic address: gesposit@na.infn.it},
Marco Figliolia$^{2}$
\thanks{Electronic address: marquarko@email.it},
Patrizia Vitale$^{2,1}$
\thanks{Electronic address: patrizia.vitale@na.infn.it}}
\affiliation{
${\ }^{1}${Istituto Nazionale di Fisica Nucleare, Sezione di
Napoli, \\
Complesso Universitario di Monte S. Angelo, \\
Via Cintia Edificio 6, 80126 Napoli, Italy \\}
${\ }^{2}${Dipartimento di
Scienze Fisiche, Complesso Universitario di Monte S. Angelo, \\
Via Cintia Edificio 6, 80126 Napoli, Italy}}

\begin{abstract}
In this paper the Seiberg--Witten map is first analyzed for
non-commutative Yang--Mills Theories with the related methods,
developed in the literature, for its explicit construction, that
hold for any gauge group. These are exploited to write down
the second-order Seiberg--Witten map for pure gravity with a
constant non-commutativity tensor. In the analysis of pure gravity
when the classical space-time solves the vacuum Einstein equations,
we find for three distinct vacuum solutions that the
corresponding non-commutative field equations do not have solution
to first order in non-commutativity, when the Seiberg--Witten map
is eventually inserted. In the attempt of understanding whether
or not this is a peculiar property of gravity,
in the second part of the paper,
the Seiberg--Witten map is considered in the simpler case of
Maxwell theory in vacuum in the absence of charges and currents.
Once more, no obvious solution of the non-commutative field
equations is found, unless the electromagnetic potential depends
in a very special way on the wave vector.
\end{abstract}

\maketitle

\section{Introduction}

Field theories on a non-commutative space can be obtained by
replacing the ordinary products with the $\star$-product. This idea
has been widely developed over the last years, mainly because the
non-commutative gauge theory turns out to be a limit of string
theory \cite{sw}. This relation with string theory makes it possible
to map non-commutative theories into commutative ones \cite{sw}
with the help of the Seiberg--Witten map which is a gauge equivalence
relation between non-commutative gauge theory and its ordinary
counterpart which is compatible with the gauge structure of the theory.
The Seiberg--Witten map for non-commutative fields can be written
for arbitrary non-Abelian gauge groups.
Explicit Seiberg--Witten maps of non-commutative fields are useful
both to understand physical predictions and to check the behaviour
of non-commutative theory itself, e.g. perturbative renormalizability.

The Seiberg--Witten maps for gauge parameter and fields can be
obtained as solutions of consistency conditions of the gauge theory
under consideration \cite{jmssw}, that are analyzed perturbatively
by expanding gauge parameter and fields in a formal power series
in the non-commutativity parameter $\theta$. This method has provided
explicit solutions for the Seiberg--Witten map of non-Abelian gauge
theories only up to the second order in
$\theta$ \cite{jmssw, gh, moller, hakikat, ana, tw}.

In the work in Ref. \cite{ulk}, the Seiberg--Witten map for gauge
parameter, gauge field and matter fields is written to all orders
for a non-commutative non-Abelian theory. The authors of Ref.
\cite{ulk} show that the map to second order for the fields
\cite{moller} can be written in terms of their counterparts to
zeroth and first order in the expansion of the map. By virtue of
this structure of the second-order calculation, in Ref. \cite{ulk}
a recursive formula has been built for all orders that satisfies
the Seiberg--Witten map.

By exploiting this correspondence, in the present paper we build
the second-order Seiberg--Witten map for a theory of pure gravity
expressed in the tetrad formalism. By doing so we introduce a set
of local Lorentz frames, whose global existence is ensured if the
classical space-time manifold is parallelizable, and we regard gravity
as a gauge theory on a non-commutative space defined by
\be
\left [x^\mu ,x^\nu \right ]_\star ={\rm i} \theta ^{\mu\nu}~,
\ee
where $\theta$ is a constant Poisson tensor, and the
$\star$-product is the associative Weyl--Moyal product
\be
\label{WMproduct}
f \star g= f\, {\rm e}^{\frac{{\rm i}}{2}\theta
^{jk}\stackrel{\leftarrow}{\partial _j}
\stackrel{\rightarrow}{\partial_k}}g.
\ee
Here we exploit the fact that the derivative of functions
satisfies the Leibniz rule with respect to the $\star$-product, i.e.
\be
\del_i(f \star g)
=(\del_i f) \star g+f \star (\del_i g),
\ee
as in the case of the ordinary product; this imples that $\theta$
is constant.

In light of all these considerations, in Sec. II we review the
fundamental concepts for building a Seiberg--Witten map for a
Yang--Mills theory, and how is it built to first order. In Sec. III,
we summarize the strategies for generalizing the Seiberg--Witten map
adopted in Refs. \cite{ulk} and \cite{cerc}.
In Sec. IV, following the work in Ref. \cite{ulk},
we apply the recursive procedure for
providing the second-order Seiberg--Witten map for the tetrad, which
can be written in terms of fields to zeroth order in the
non-commutativity parameter, and in terms of first-order solutions.
Section V writes the pure-gravity action to second
order in non-commutativity.
In Sec. VI we consider pure gravity when the classical
space-time solves the vacuum Einstein equations
(three distinct cases). The resulting
non-commutative field equations are found not to have solution
to first order in non-commutativity, when the Seiberg--Witten
map is eventually inserted. In the attempt of understanding
whether or not this is a peculiar property of gravity, Sec. VII
studies the simpler case of vacuum Maxwell theory in the absence
of charges and currents. Concluding remarks
and open problems are presented in Sec. VIII, while the Appendix
describes the attempt of solving the non-commutative field equations
of pure gravity without using the Seiberg--Witten map.

\section{The Seiberg--Witten map to first order}

For a Yang--Mills theory, the gauge potential is a
Lie-algebra-valued 1-form. Following the notation in Ref.
\cite{sw}, the gauge transformations
for components of potential and field strength read as
\be
\d_{\L} A_{\mu} = \del_{\mu}
\L - {\rm i}[A_{\mu} , \L] \equiv D_{\mu}\L , \quad \d_{\L}
{F}_{\mu\nu} = {\rm i} [\L , {F}_{\mu\nu}],
\ee
where the Lie-algebra indices (frequently written upstairs
and taken from the beginning of the Greek alphabet)
are omitted for simplicity.
With this understanding, one can write the Yang--Mills
field strength in the form
\be
\label{F} {F}_{\mu\nu} =
\del_\mu A_{\nu} - \del_\nu A_{\mu} - {\rm i}[A_{\mu} , A_{\nu}],
\ee
out of which one can get the gauge curvature 2-form
\begin{equation}
F=F_{\mu \nu}{\rm d}x^{\mu} \wedge {\rm d}x^{\nu}.
\end{equation}
For a non-commutative Yang--Mills theory, one uses the same formulae
for gauge transformation and field strength, with the understanding
that the ordinary product of matrices is replaced by the Moyal
$\star$-product \cite{sw}, i.e.
\be
\hd_{\hL} \hA_{\mu} = \del_{\mu} \hL - {\rm i}[\hA_{\mu} ,
\hL]_{\star} \equiv \hD_{\mu}\hL , \quad \hd_{\hL}
\hat{F}_{\mu\nu} = {\rm i} [\hL , \hat{F}_{\mu\nu}]_{\star},
\label{gauge}
\ee
where
\be
\label{hF}
\hat{F}_{\mu\nu} = \del_\mu \hA_{\nu} -
\del_\nu \hA_{\mu} - {\rm i}[\hA_{\mu} , \hA_{\nu}]_{\star}
\label{trasfg}
\ee
is the non-commutative field strength of the non-commutative
gauge potential $\hA$. The resulting theory reduces to the
familiar Yang--Mills theory with group $U(N)$ as $\theta\to 0$.

The Seiberg--Witten map relates non-commutative fields 
and their commutative counterparts through a gauge equivalence
relation \cite{sw}:
\be
\label{swe}
\hA_\mu (A;\theta) + \hd_{\hL}\hA_\mu
(A;\theta) = \hA_\mu (A+\d_{\L}A;\theta ) ,
\ee
where $A$ and $\L$ are the commutative gauge field (or potential) and
gauge parameter, respectively.

Equation (\ref{swe}) can be re-written in the form
\be
\label{swe2}
\hd_{\hL}\hA_\mu (A,\theta) = \hA_\mu (A+\d_{\L}A;\theta ) -
\hA_\mu (A;\theta) = \d_\L \hA_{\mu}(A;\theta) .
\ee
To first order in $\theta$, denoted by the superscript $1$,
the formulae (\ref{gauge}) become
\begin{equation}
\delta_\L A^1_i[a]=A^1_\L(A+\d_\L A)-A^1_\L(A)
=\partial_i\L^1[A]+{\rm i}[\L^1[A],A_i]+{\rm i}[\L,A^1_i[a]]
-\frac{1}{2}\theta^{kl}\{\partial_k\L,
\partial_lA_i\}.
\label{eqswI}
\end{equation}
The solution of Eq. (\ref{eqswI}) to first order in
$\theta$ is (see Eq. (3.5) of \cite{sw})
\be
\label{a11}
A_{\g}^{1} = -\frac{1}{4}\theta^{\k\l} \{ A_\k
, \partial_{\lambda} A_{\gamma} + F_{\l\g} \} ,\quad
 \L^{1} = \frac{1}{4}\theta^{\k\l} \{ \del_{\k}\L,A_\l \}.
\ee
We can check it in particular for the gauge field, and analogous
procedure holds for the gauge parameter and field strength.
By applying the gauge transformation (\ref{gauge})
to the first of Eqs. (\ref{a11}), and applying
again Eq. (\ref{gauge}) within it, one finds
\begin{eqnarray}
\delta_\L A^1_\g[a]&=& -\frac{1}{4}\theta^{\k\l} \{ \d_\L A_\k
,\del_\l A_\g + F_{\l\g} \} -\frac{1}{4}\theta^{\k\l} \{ A_\k
,\del_\l \d_\L A_\g + \d_\L F_{\l\g} \}
\nonumber \\
&-& \frac{1}{4}\theta^{\k\l} \{ \del_{\k} \L
- {\rm i}[A_{\k} , \L]
,\del_\l A_\g + F_{\l\g} \} -\frac{1}{4}\theta^{\k\l} \{ A_\k
,\del_\l \del_{\g} \L\}
\nonumber \\
&+&\frac{{\rm i}}{4}\theta^{\k\l} \{ A_\k ,
[\del_\l A_{\g} , \L] \}
+\frac{{\rm i}}{4}\theta^{\k\l}
\{ A_\k , [A_{\g} , \del_\l \L] \}
\nonumber \\
&-& \frac{{\rm i}}{4}\theta^{\k\l} \{ A_\k , [\L ,
{F}_{\l\g}]\}.
\label{eqsw2}
\end{eqnarray}
On the other hand, by inserting Eqs. (\ref{a11}) into the right-hand
side of Eq. (\ref{eqswI}), one finds
\begin{eqnarray}
\delta_\L A^1_\g[a]&=& \frac{1}{4}\theta^{\k\l} \{
\del_{\g}\del_{\k}\L, A_\l \} + \frac{1}{4}\theta^{\k\l} \{
\del_{\k}\L, \del_{\g}A_\l \}
+ \frac{{\rm i}}{4}\theta^{\k\l} [\{
\del_{\k}\L, A_\l \}, A_\g] \
\nonumber \\
&-& \frac{{\rm i}}{4}\theta^{\k\l} [\L, \{A_\k,
\del_{\l}A_\g + F_{\l\g}\}]
-\frac{1}{2}\theta^{\k\l} \{\del_{\k}\L,
\del_{\l}A_\g\}.
\label{eqsw3}
\end{eqnarray}
By comparing Eq. (\ref{eqsw2}) with Eq. (\ref{eqsw3}), and using
the following property:
\begin{equation}
\{A, [B,C]\} + \{B, [A,C]\}= [\{A, B\},C],
\end{equation}
one checks the equality and hence Eq.  (\ref{eqswI}) is solved by
(\ref{a11}).

\section{The Seiberg--Witten map to all orders }

In this section we are going to summarize the various ways
available for building a Seiberg--Witten map to all orders in
$\theta$, in particular those proposed in
Refs. \cite{ulk}, \cite{wulk}, \cite{cerc}.
\vskip 0.3cm
\noindent
{\it First way}.

In Ref. \cite{ulk} the authors consider, in order to build a
Seiberg--Witten map to all orders, the following series expansion
in $\theta$:
\bea
\hL_\a &=& \a + \L_\a^{1} + \cdots + \L_\a^{n} + \cdots,\\
\hA_\mu &=& A_\mu + A_{\mu}^{1} + \cdots + A_{\mu}^{n} + \cdots,
\nonumber
\eea
where the zeroth order terms $\a$ and $A_\mu$ are the commutative
counterparts of $\hL_\a$ and $\hA_\mu$. Moreover, one denotes by
${\star}^r$ the $r$-th order term in the expansion of the
$\star$-product, i.e.
\begin{equation}
f(x){\star}^{r}g(x) = \frac{1}{r!}\left(\frac{{\rm i}}{2}\right) ^r
\theta^{\mu_1\nu_1} \cdots \theta^{\mu_r\nu_r}
\del_{\mu_1}\cdots\del_{\mu_r}f(x)
\del_{\nu_1}\cdots\del_{\nu_r}g(x) .
\end{equation}
By exploiting the first-order Seiberg--Witten map given in (2.9), 
one can obtain its second-order form from the Eqs. \cite{ulk}
\be
\L_\a^{2} = -\frac{1}{8}\theta^{\k\l}\left(
\{A_{\k}^{1}, \del_{\l}\a \} + \{A_{\k}, \del_{\l}\L_\a^{1}\}
\right) -\frac{{\rm i}}{16}\theta^{\k\l}\theta^{\mu\nu}[\del_\mu A_\k ,
\del_\nu \del_{\l}\a ], 
\label{l2}
\ee
\bea
\label{a22}
A_{\g}^{2}
&=& -\frac{1}{8}\theta^{\k\l}\left( \{A_{\k}^{1}, \del_\l A_\g +
F_{\l\g} \}
+ \{A_{\k}, \del_\l A_\g ^1 + F_{\l\g} ^{1}\} \right)
\nonumber \\
&-& \frac{{\rm i}}{16}\theta^{\k\l}\theta^{\mu\nu}[\del_\mu A_\k ,
\del_\nu (\del_\l A_\g + F_{\l\g} ) ], 
\eea
and the authors of Ref. \cite{ulk} stress that terms linear in $\theta$
in the formulae above result from an expansion where the
first-order terms are given by (2.9), and hence quadratic
terms are related to the expansion of the $\star$-product in powers
of $\theta$. Thus, by inspection of the structure of the two
expansions to first and second order, respectively, the authors
\cite{ulk} make the following conjecture for the general structure
of the Seiberg--Witten map to all orders in $\theta$, i.e. a
recursive formula to all orders reading as
\be
\label{nl}
\L^{n+1} _\a =
-\frac{1}{4(n+1)}\theta^{\k\l}\sum_{{p+q+r=n}} \{A_\k ^p , \del_\l
\L^q _\a \}_{{\star}^r},
\ee
\be
\label{na}
A_{\g}^{n+1} =
-\frac{1}{4(n+1)}\theta^{\k\l}\sum_{{p+q+r=n}} \{ A_{\k}^{p},
\del_\l A^q_\g + F^{q}_{\l\g} \}_{{\star}^{r}},
\ee
where the first-order solution is obtained by setting $n=0$,
the second-order solution by setting $n=1$, and so on.
\vskip 0.3cm
\noindent
{\it Second way}.

There exists an alternative approach to studying the Seiberg--Witten
map, that relies upon finding the same solution starting from a
differential equation introduced in \cite{sw}, whose solution
is given by
\be
\label{wln}
\L^{n+1} _\a =
-\frac{1}{4(n+1)!}\theta^{\mu\nu}\theta^{\mu_1\nu_1} \cdots
\theta^{\mu_n\nu_n} \left( \frac{\del^{n}}{\del\theta^{\mu_1\nu_1}
\cdots \del\theta^{\mu_n\nu_n} } \sum_{p+q+r=n} \{ A_{\mu}^{p},
\del_\nu \L^{q} _\a  \}_{{\star}^{r}} \right)_{\theta =0},
\ee
\bea
\label{wlm}
A^{n+1}_\g &=&
-\frac{1}{4(n+1)!}\theta^{\mu\nu}\theta^{\mu_1\nu_1} \cdots
\theta^{\mu_n\nu_n} \left( \frac{\del^{n}}{\del\theta^{\mu_1\nu_1}
\cdots \del\theta^{\mu_n\nu_n} } \{\hA^{(n)}_{\mu_1},\del_{\nu_1}
\hA^{(n)} _\g + \hat{F}^{(n)}_{\nu_1 \g}  \}_{\star}
\right)_{\theta =0}.
\eea
Equations (\ref{wln}), (\ref{wlm}) reduce to the previous
recursive formulae (\ref{nl}), (\ref{na}).
\vskip 0.3cm
\noindent
{\it Third way}

Yet another approach to studying the Seiberg--Witten map relies upon
a differential equation of time-evolution type \cite{cerc}. The
authors of this paper introduce a time parameter $t$ such that
$\theta^{ij} \rightarrow t \,
\theta^{ij}$, $\L \rightarrow \L(t)$ and $A_i \rightarrow A_i(t)$.
In such a way, $\L$ and  $A_i$ acquire a time dependence through
$\theta$. The evolution equation provides a method useful for the
evaluation of higher-order terms by differentiation with respect
to $t$. One can then compute $A_i^{(n)}$ and $\L^{(n)}$ according to
\be
\L^{(n)}=\frac{1}{n!} {\del^n \L(t) \over \del
t^n} \mid_{t=0}, \qquad A_i^{(n)}=\frac{1}{n!} {\del^n A_i(t)
\over \del t^n} \mid_{t=0}.
\ee

There exist other techniques for finding the Seiberg--Witten map
at higher orders, based upon the homotopy operator, applied order
by order, as suggested in Refs. \cite{BCPVZ,BCZ}.

\section{Seiberg--Witten map for pure gravity}

In this section we relate a non-commutative version of pure gravity
to its commutative counterpart by using the Seiberg--Witten map,
relying upon the work in Ref. \cite{miao10}. In this case the
Seiberg--Witten map relates the non-commutative degrees of
freedom $\hat{\omega}_\mu$, $\hat{e}_\mu$ and $\hat{\L}$ to their
commutative counterparts, i.e. classical spin connection
$\omega_\mu$, tetrad $e_\mu$ and gauge parameter $\L$.

The map between $\hat{\omega}_\mu$ and $\omega_\mu$,
and between $\hat{e}_\mu$ and ${e}_\mu$
to first order in the constant non-commutativity $\theta^{\mu\nu}$
is given by \cite{miao10}
\begin{eqnarray}
\hat{e}_\mu = e_\mu - \frac{1}{2} \th^{\l\s} \{\om_\l, \p_\s e_\mu
+ \frac{{\rm i}}{2} [e_\mu, \om_\s]\},
\label{eSWMap}
\end{eqnarray}
\begin{equation}
\hat{\omega}_\mu = \omega_\mu - \frac{1}{4} \th^{\l\s}\{\om_\l,
\p_\s \om_\mu + R_{\s\mu}\}. \label{OmMap}
\end{equation}
We now perform in the first way, as seen in the previous Section, the
construction of the Seiberg--Witten map for pure gravity with the
understanding that
$A_{\k}^{p}=\om_{\k}^{p}$ and  $F^{q}_{\l\g}= R^{q}_{\l\g}$:
\begin{eqnarray}
\hat{e}_\mu^{n+1} = - \frac{1}{2(n+1)} \th^{k\l}
\left(\sum_{{p+q+r=n}} \{ \om_{\k}^{p},
\del_\l e^q_\mu \}_{{\star}^{r}}
+ \frac{{\rm i}}{2}\sum_{p+q+r+s=n}\{\om_k^p,  [e_\mu^q,
\om_\l^s]\}_{{\star}^{r}}\right)
\label{eSWMapgrav1}.
\end{eqnarray}
In particular, for the second-order map for the spin-connection
one now finds (this order implies setting $n=1$)
\bea
\label{omgrav}
\om_{\g}^{2} &=&
-\frac{1}{8}\theta^{\k\l}\left( \{\om_{\k}^{1}, \del_\l \om_\g +
R_{\l\g} \}
+ \{\om_{\k}, \del_\l \om_\g ^1 + R_{\l\g} ^{1}\} \right) \\
&& -\frac{{\rm i}}{16}
\theta^{\k\l}\theta^{\mu\nu}[\del_\mu \om_\k ,
\del_\nu (\del_\l \om_\g + R_{\l\g} ) ]
\nonumber
\eea
while for the tetrad
\begin{eqnarray}
e_\mu^{2}& = &- \frac{1}{4} \theta^{\k\l} \left(\{ \om_{\k}^{1},
\del_\l e_\mu \}+ \{ \om_{\k}, \del_\l e^1_\mu \}\right) -
\frac{{\rm i}}{8}\theta^{k\l}\theta^{\s\nu}
[\p_\s \omega_k,\p_\nu\p_\l e_\mu]
\nonumber \\
&& -\frac{{\rm i}}{8}\theta^{k\l}\left( \{\om_k^1,
[e_\mu, \om_\l]\}+\{\om_k, [e_\mu^1, \om_\l]\}+ \{\om_\l, [e_\mu,
\om_\l^1]\}\right)\nonumber \\
&& +\frac{1}{16}\theta^{k\l}\theta^{\s\nu}
\left([\p_\s\om_k, [\p_\nu
e_\mu, \om_\l]]+[\p_\s\om_k, [e_\mu,\p_\nu \om_\l]]+ \{\om_k,
\{\p_\s e_\mu,\p_\nu \om_\l\}\}\right).
\label{eSWMapgrav2}
\end{eqnarray}

From now on, with our notation, $V$ is the tetrad 1-form
expanded on the basis of Dirac $\gamma$-matrices, i.e.
\begin{equation}
V=V_\mu {\rm d}x^\mu=(V_{\mu}^{a}\gamma_{a}
+{\widetilde V}_{\mu}^{a}\gamma_{a}\gamma_{5})
{\rm d}x^\mu=\hat{e}_\mu {\rm d}x^\mu=
(\hat{e}^{(0) a}_\mu \g_a + \hat{e}^{(1) a}_{\mu 5} \g_5
\g_a){\rm d}x^\mu,
\label{(2)}
\end{equation}
while $R$ is the curvature 2-form
\begin{equation}
R=d\Omega-\Omega \wedge_{*} \Omega.
\label{(3)}
\end{equation}
Our convention for the Minkowski metric is
$\eta_{ab}={\rm diag}(1,-1,-1,-1)$, and we define
$\sigma_{ab} \equiv - \frac{i}{4} [\g_a, \g_b]=-\frac{i}{2}
\gamma_{ab}$. One thus finds
\begin{equation}
\hat{\omega}_\mu = \frac{1}{2} \hat{\omega}^{ ab}_\mu \s_{ab} +
\hat{a}_\mu + {\rm i} \hat{b}_{\mu 5} \g_5,
\label{1}
\end{equation}
\begin{equation}
\hat{R}_{\rho\s} = \frac{{\rm i}}{2} \hat{R}_{\rho\s}^{ab} \s_{ab}
+ {\rm i} \hat{r}_{\rho\s} +  \hat{\rtilde}_{\rho\s } \g_5.
\end{equation}
Hereafter we use the following properties of $\gamma$-matrices:
\begin{eqnarray}
 & &\gamma_5 \sigma_{ab} = \sigma_{ab}\gamma_5
={{\rm i} \over 2} \epsilon_{abcd} \sigma^{cd},
 \\
  & &\{\sigma_{ab}, \gamma_5\} = {\rm i}
\epsilon_{abcd} \sigma^{cd},
\\
  & &[\sigma_{ab}, \gamma_5] = 0,
 \\
 & & [\sigma_{ab}, \gamma_c] =
{\rm i} (\eta_{ac} \g_b - \eta_{bc} \gamma_a), \\
& &\{\sigma_{ab}, \gamma_c\} = - \epsilon_{abc}^{~~~d}\gamma_5\gamma_d, \\
& &\{\sigma_{ab}, \sigma_{cd}\} = \frac{1}{2} ({\rm i} \epsilon_{abcd}
\gamma_5 + \eta_{ac} \eta_{bd} - \eta_{ad} \eta_{bc}),
\end{eqnarray}
from which
\begin{eqnarray}
&&[\s_{ab},\s_{cd}]={\rm i}(\eta_{ac}\sigma_{bd}
-\eta_{bc}\sigma_{ad}-\eta_{ad}\sigma_{bc}+\eta_{bd}\sigma_{ac}),
\\
 & & [\gamma_5\g_e, \s_{ab}] = {\rm i} \g_5(\eta_{be}
\gamma_a-\eta_{ae} \g_b ) , \\
& &\{[\g_5\g_e, \s_{ab}],\s_{cd}\}={\rm i}(\eta_{ae}
\ep_{cdb}^{~~~f}-\eta_{be}\ep_{cda}^{~~~f} )\g_f, \\
& &\{[\g_c, \g_5],\s_{ab}\}=2\{\g_c \g_5,\s_{ab}\} =
2\ep_{abc}^{~~~d}\gamma_{d}, \\
& & \{[\g_c, \s_{ab}],\g_5\}=0,\\ &&
\{\s_{ab},\g_c\g_5\}=\ep_{abc}^{~~~d}\g_d ,
\\ &&
\{\s_{ab},\{\g_c, \s_{ef}\}\}=-\ep_{efc}^{~~~d}\ep_{abmd}\g_{m} ,
\\ && [\s_{ab},[\g_c, \s_{ef}]]=(\eta_{ec}\eta_{af} -
\eta_{fc}\eta_{ae})\gamma_b+(\eta_{fc}\eta_{be} -
\eta_{ec}\eta_{bf})\gamma_{a},
\end{eqnarray}
and hence, by using (\ref{eSWMapgrav2}) and (\ref{1})
\begin{eqnarray}
e_\mu^{2}& = &- \frac{1}{4} \theta^{\k\l} \left(\{ \om_{\k}^{1},
\del_\l e_\mu \}+ \{ \om_{\k}, \del_\l e^1_\mu \}\right) -
\frac{{\rm i}}{8}\theta^{\k\l}\theta^{\s\nu}
[\p_\s \omega_\k,\p_\nu\p_\l e_\mu]
\nonumber \\
&& -\frac{{\rm i}}{8}\theta^{\k\l}\left( \{\om_\k^1,
[e_\mu, \om_\l]\}+\{\om_\k, [e_\mu^1, \om_\l]+ \{\om_\k, [e_\mu,
\om_\l^1]\}\}\right)\nonumber \\&&
+\frac{1}{16}\theta^{\k\l}\theta^{\s\nu}\left([\p_\s\om_\k,
[\p_\nu e_\mu, \om_\l]]+[\p_\s\om_\k, [e_\mu,\p_\nu \om_\l]]+
\{\om_\k, \{\p_\s e_\mu,\p_\nu \om_\l\}\}\right),
\label{eSWMapgrav2svil}
\end{eqnarray}
where $\om_{\mu}^{1}= \hat{a}^{(1)}_\mu + {\rm i} \hat{b}^{(1)}_{\mu 5}
\g_5$, $e_\mu^1= \hat{e}^{(1) a}_{\mu 5} \g_5 \g_a$, while
$\om_{\mu}=\frac{1}{2} \hat{\omega}^{(0) ab}_\mu \s_{ab}$, $e_\mu=
\hat{e}^{(0) a}_\mu \g_a $. Thus, by substitution, a careful
calculation leads eventually to (cf. Ref. \cite{AC12b})
\begin{eqnarray}
e_\mu^{2}& = &- \frac{1}{4} \theta^{\k\l} \left( 2\hat{a}^{(1)}_\k
\del_\l \hat{e}^{(0) a}_\mu\g_a -\frac{1}{2} \hat{\omega}^{(0)
ab}_\k  \del_\l \hat{e}^{(1) c}_{\mu 5}\ep_{abc}^{~~~d}\g_d\right)
\nonumber \\
&&+\frac{1}{16}\theta^{\k\l}\theta^{\s\nu}\p_\s  \hat{\omega}^{(0) ab}_\k
\p_\nu\p_\l \hat{e}^{(0) c}_\mu(\eta_{ac} \g_b - \eta_{bc}
\gamma_a)
\nonumber \\
&& -\frac{1}{8}\theta^{\k\l}\left(\hat{a}^{(1)}_\k\hat{e}^{(0)
c}_\mu
 \hat{\omega}^{(0) ab}_\l (\eta_{ac} \g_b - \eta_{bc} \gamma_a)
\right.
\nonumber \\
&&-\frac{1}{4}
\hat{\omega}^{(0) ab}_\k\hat{e}^{(1) c}_{\mu 5} \hat{\omega}^{(0)
de}_\l (\eta_{dc}
\ep_{abe}^{~~~f}-\eta_{ec}\ep_{abd}^{~~~f} )\g_f
\nonumber \\
&&\left.  -\hat{\omega}^{(0) ab}_\k\hat{e}^{(0) c}_\mu
\hat{b}^{(1)}_{\l 5} \ep_{abc}^{~~~d}\gamma_{d}\right)
\nonumber \\
&&+\frac{1}{64}\theta^{\k\l}\theta^{\rho\nu}\times
\biggr[\left(\p_\rho \hat{\omega}^{(0) ab}_\k\p_\nu \hat{e}^{(0)
c}_\mu \hat{\omega}^{(0) de}_\l+ \p_\rho \hat{\omega}^{(0) ab}_\k
\hat{e}^{(0) c}_\mu\p_\nu \hat{\omega}^{(0)
de}_\l\right)
\nonumber \\
&\times& ((\eta_{ec}\eta_{af} -
\eta_{fc}\eta_{ae})\gamma_b+(\eta_{fc}\eta_{be} -
\eta_{ec}\eta_{bf})\gamma_a)
\nonumber \\
&+&  \hat{\omega}^{(0) ab}_\k\p_\rho
\hat{e}^{(0) c}_\mu \p_\nu\hat{\omega}^{(0) de}_\l
\ep_{dec}^{~~~d}\ep_{abd}^{~~~m}\g_m\biggr].
\label{eSWMapgrav2svil3}
\end{eqnarray}
To second order in $\theta^{\k\l}$, the resulting components
along $\g_a$ and $\g_5\g_a$ are
\begin{eqnarray}
\hat{e}^{(2)h}_\mu& = &- \frac{1}{4} \theta^{\k\l} \left(
2\hat{a}^{(1)}_\k \del_\l \hat{e}^{(0) h}_\mu -\frac{1}{2}
\hat{\omega}^{(0) ab}_\k  \del_\l \hat{e}^{(1) c}_{\mu
5}\ep_{abc}^{~~~h}\right)  \nonumber \\
&& -\frac{1}{4}\theta^{\k\l}\left(\hat{a}^{(1)}_\k\hat{e}^{(0)
c}_\mu
 \hat{\omega}^{(0) ah}_\l \eta_{ac}
\right.-\frac{1}{4}\hat{\omega}^{(0) ab}_\k\hat{e}^{(1) c}_{\mu 5}
\hat{\omega}^{(0) de}_\l \eta_{dc} \ep_{abe}^{~~~h}
\nonumber \\
&&\left.  -\frac{1}{2}\hat{\omega}^{(0) ab}_\k\hat{e}^{(0) c}_\mu
\hat{b}^{(1)}_{\l 5} \ep_{abc}^{~~~h}\right)
\nonumber \\
&&+\frac{1}{32}\theta^{\k\l}\theta^{\s\nu}\times \left[\left(\p_\s
\hat{\omega}^{(0) ah}_\k\p_\nu \hat{e}^{(0) c}_\mu
\hat{\omega}^{(0) ef}_\l+ \p_\s \hat{\omega}^{(0) ah}_\k
\hat{e}^{(0) c}_\mu\p_\nu \hat{\omega}^{(0)
ef}_\l\right)\right.
\nonumber \\
&& \times(\eta_{ec}\eta_{af} -
\eta_{fc}\eta_{ae}) +  \frac{1}{2}\hat{\omega}^{(0) ab}_\k\p_\nu
\hat{e}^{(0) c}_\mu \p_\s\hat{\omega}^{(0) ef}_\l
\ep_{efc}^{~~~d}\ep_{abd}^h \nonumber \\ && \left.+4\p_\s
\hat{\omega}^{(0) ah}_\k \p_\nu\p_\l \hat{e}^{(0) c}_\mu\eta_{ac}
\right],
\\
\hat{e}^{(2)a}_{\mu5} & =& 0. \label{eSWMapgrav2svil4}
\end{eqnarray}
respectively.

As far as the spin-connection is concerned, one has
(recall that $\frac{1}{2}
\hat{R}^{(1)}_{\l\g}={\rm i}r, \frac{{\rm i}}{2} \hat{R}^{(1)}_{\l\g
5}=\tilde{r}$ )
\bea
\label{omgrav}
\om_{\g}^{2} &=&
-\frac{1}{8}\theta^{\k\l}\left( \{\om_{\k}^{1}, \del_\l \om_\g +
R_{\l\g} \}
+ \{\om_{\k}, \del_\l \om_\g ^1 + R_{\l\g} ^{1}\} \right) \\
&& -\frac{{\rm i}}{16}\theta^{\k\l}\theta^{\mu\nu}[\del_\mu \om_\k ,
\del_\nu (\del_\l \om_\g + R_{\l\g} ) ] \nonumber
\eea
from which, by insertion of the various terms, we find
(cf. Ref. \cite{AC12b})
\bea
\label{omgrav2}
\om_{\g}^{2} &=&
-\frac{1}{8}\theta^{\k\l}\left(
2\hat{a}^{(1)}_\k(\frac{1}{2}\del_\l  \hat{\omega}^{(0)
ab}_\g+\frac{{\rm i}}{2} \hat{R}_{\l\g}^{(0)ab})\s_{ab} -\frac{1}{2}
\hat{b}^{(1)}_{\k5}(\del_\l  \hat{\omega}^{(0) ab}_\g +
\hat{R}_{\l\g}^{(0)ab})   \epsilon_{abcd} \sigma^{cd}
 \right.
 \nonumber \\
&&\left.
+\hat{\omega}^{(0) ab}_\k(\del_\l \hat{a}^{(1)}_\g +{\rm i}
\hat{r}^{(1)}_{\l\g})\s_{ab}+ \frac{1}{2}\hat{\omega}^{(0)
ab}_\k({\rm i}\del_\l\hat{b}^{(1)}_{\g 5} +
\hat{\tilde{r}}^{(1)}_{\l\g }) \epsilon_{abcd} \sigma^{cd}\right)
\nonumber \\
&&+\frac{1}{64}\theta^{\k\l}\theta^{\mu\nu}\del_\mu
\hat{\omega}^{(0) ab}_\k\del_\nu (\del_\l \hat{\omega}^{(0)
cd}_\g+\frac{{\rm i}}{2} \hat{R}_{\l\g}^{(0)cd})
(\eta_{ac}\sigma_{bd}
-\eta_{bc}\sigma_{ad}-\eta_{ad}\sigma_{bc}+\eta_{bd}\sigma_{ac}),
\eea
from which the components along
$\s_{ab}$, $I$, $\g_5$ read as
\begin{eqnarray}
&&\hat{\omega}^{(2) ab}_\g=-\frac{1}{8}\theta^{\k\l}\left(
2\hat{a}^{(1)}_\k(\frac{1}{2}\del_\l  \hat{\omega}^{(0) ab}_\g+
\hat{R}_{\l\g}^{(0)ab}) -\frac{1}{2} \hat{b}^{(1)}_{\k5}(\del_\l
\hat{\omega}^{(0) cd}_\g + {\rm i} \hat{R}_{\l\g}^{(0)cd})
\epsilon_{abcd} \right.
\nonumber \\
&&\left.
+\hat{\omega}^{(0) ab}_\k(\del_\l \hat{a}^{(1)}_\g +{\rm i}
\hat{r}^{(1)}_{\l\g})+ \frac{1}{2}\hat{\omega}^{(0)
cd}_\k({\rm i}\del_\l\hat{b}^{(1)}_{\g 5} +
\hat{\tilde{r}}^{(1)}_{\l\g }) \epsilon_{abcd} \right)
\nonumber \\
&&+\frac{1}{64}\theta^{\k\l}\theta^{\mu\nu}\del_\mu
\hat{\omega}^{(0) ef}_\k\del_\nu (\del_\l \hat{\omega}^{(0)
cd}_\g+\frac{{\rm i}}{2} \hat{R}_{\l\g}^{(0)cd})
(\eta_{ec}\delta_f^a\delta_d^b
-\eta_{fc}\delta_e^a\delta_d^b-\eta_{ed}\delta_f^a\delta_c^b+\eta_{fd}\delta_e^a\delta_c^b)\\
&&
\hat{a}^{(2)}_\g=\omega^{(2)}_\g=0, \\
&& \hat{b}^{(2)}_{\g 5}=\tilde{\omega}^{(2)}_\g=0.
\end{eqnarray}
Since the second-order part of the $\westar$ product of 1-forms
is skew-symmetric, one has in our case
\begin{eqnarray}
& & R^{(2)ab}= d \om^{(2)ab}  +\om^{(0)b}_c \we_{\star II}\om^{(0)ca}
\nonumber \\
&& + \om^{(2)b}_c \we\om^{(0)ca}
+\om^{(0)b}_c \we\om^{(2)ca}
+2{\rm i} \om^{(0)ab}
\we_{\star I} \om^{(1)}   -  \epsilon^{ab}_{~~cd}
\om^{(0)cd} \we_{\star I} \omtilde^{(1)}, \\
& &   r^{(2)} = 0,
\nonumber \\
& &  \rtilde^{(2)} = 0.
\end{eqnarray}
A neater geometric derivation of the second-order Seiberg--Witten
map for pure gravity can be found in Ref. \cite{July2012}, with
emphasis on the Seiberg--Witten differential equation for the
action itself.

\section{Second-order action functional}

The Aschieri--Castellani action for pure gravity
\cite{July2012,AC09, AC11,AC12} reads as
\begin{eqnarray}
S= & &  \int R^{ab} \westar (\hat{e}^{c}
\westar \hat{e}^{d} - \hat{e}^{c}_5 \westar \hat{e}^{d}_5 )
\epsilon_{abcd}
\nonumber \\
& &  + 2{\rm i}~ R^{ab} \westar (-\hat{e}_{a}
\westar  \hat{e}_{b5}  + \hat{e}_{a5}  \westar
\hat{e}_{b})
\nonumber \\
& &+ 4{\rm i}~r \westar(\hat{e}^{a} \westar \hat{e}_{a5}
- \hat{e}^{a}_5  \westar
\hat{e}_{a})
\nonumber \\
& & + 4{\rm i}~ \rtilde \westar (\hat{e}^{a}
\westar \hat{e}_{a}-\hat{e}^{a}_5 \westar \hat{e}_{a5}).
\nonumber \\
\end{eqnarray}
Bearing in mind that
$r^{(0)}=\rtilde^{(0)}=R^{ab(1)}=\hat{e}^{(1)}=\hat{e}^{(0)c}_5=0$,
its form to second order in $\theta$ reduces to
\begin{eqnarray}
 S^{(2)} = & &  \int [R^{ab(0)} \westar \hat{e}^{(0)c}
\westar \hat{e}^{(0)d}+
R^{ab(0)} \westar (\hat{e}^{(0)c} \we \hat{e}^{(2)d}
+\hat{e}^{(2)c} \we \hat{e}^{(0)d}-
\hat{e}^{(1)c}_5 \we \hat{e}^{(1)d}_5)
\nonumber \\
&& + R^{ab(2)} \westar \hat{e}^{(0)c} \westar \hat{e}^{(0)d}]
 \epsilon_{abcd}
\nonumber \\
 & &  + 2{\rm i}~ R^{ab (0)} \westar (
-\hat{e}_{a}^{(0)} \westar  \hat{e}_{b5}^{(1)}
+ \hat{e}_{a5}^{(1)} \westar \hat{e}_{b}^{(0)})
\nonumber \\
& &+ 4{\rm i}~r^{(1)} \westar(\hat{e}^{(0)a}  \westar
\hat{e}^{(1)}_{a5}  - \hat{e}^{(1)a}_5 \westar
\hat{e}^{(0)a})
\nonumber \\
& & + 4{\rm i}~ \rtilde^{(1)} \westar(\hat{e}^{a}
\westar \hat{e}_{a}-\hat{e}^{a}_5 \westar \hat{e}_{a5}).
\nonumber \\
\end{eqnarray}
Since the $\westar$ product of 1-forms has a first-order part in
$\theta$ which is symmetric, while its second-order term is
skew-symmetric, one finds
\begin{eqnarray}
S^{(2)} &=&  \int \Bigr[R^{ab(0)} \we_{\star II}\hat{e}^{(0)c}
\we \hat{e}^{(0)d}+
R^{ab(0)} \we\hat{e}^{(0)c} \we_{\star II} \hat{e}^{(0)d} +
R^{ab(0)} \we(\hat{e}^{(0)c} \we_{\star II} \hat{e}^{(0)d}
\nonumber \\
&+& \hat{e}^{(0)c} \we \hat{e}^{(2)d}+\hat{e}^{(2)c} \we \hat{e}^{(0)d}-
 \hat{e}^{(1)c}_5 \we \hat{e}^{(1)d}_5)
\nonumber \\
&+& R^{ab(2)} \we \hat{e}^{(0)c} \we \hat{e}^{(0)d}]
\epsilon_{abcd}-4{\rm i}~ R^{ab (0)} \we \hat{e}_{a}^{(0)}
\we_{\star I} \hat{e}_{b5}^{(1)}
\nonumber \\
&+& 8{\rm i}~r^{(1)} \we\hat{e}^{(0)a}  \we  \hat{e}^{(1)}_{a5}
+4{\rm i}~ \rtilde^{(1)} \we\hat{e}^{a} \we_{\star I} \hat{e}_{a}.
\end{eqnarray}
Variation of the action with respect to the components of the
tetrad, i.e. $\hat V^a$ and  ${\widetilde {\hat V}}{}^a$,
yields the non-commutative field equations first obtained
in Ref. \cite{asch2011}
\begin{eqnarray} &\;&
-\Bigr(\hat V^{d}\wedge_{*}\hat R^{ab}+\hat R^{ab}\wedge_{*}\hat
V^{d}\Bigr)
\varepsilon_{abcd}+{\rm i}(\eta_{bc}\eta_{ad}-\eta_{ac}\eta_{bd})
\Bigr(\hat R^{ab}\wedge_{*}{\widetilde {\hat V}}{}^{d}
-{\widetilde {\hat V}}{}^{d}
\wedge_{*}\hat R^{ab}\Bigr)
\nonumber \\
& &~~+ 4{\rm i} \eta_{dc}\Bigr({\widetilde {\hat r}}\wedge_{*}\hat V^{d}
-\hat V^{d}\wedge_{*}{\widetilde {\hat r}}\Bigr) +4
\eta_{dc}\Bigr({\widetilde {\hat V}}{}^{d}\wedge_{*}{\hat r}
+{\hat r} \wedge_{*}{\widetilde {\hat V}}{}^{d}\Bigr)=0,
\label{(9)}
\end{eqnarray}
\begin{eqnarray}
&\;& -\Bigr({\widetilde {\hat V}}{}^{d}\wedge_{*}\hat R^{ab} +\hat
R^{ab}\wedge_{*}{\widetilde {\hat V}}{}^{d}\Bigr)
\varepsilon_{abcd}+{\rm i}(\eta_{bc}\eta_{ad}-\eta_{ac}\eta_{bd})
\Bigr(\hat R^{ab}\wedge_{*}\hat V^{d}-\hat V^{d}
\wedge_{*}\hat R^{ab}\Bigr)
\nonumber \\
& &~~+ 4{\rm i} \eta_{dc}\Bigr({\widetilde {\hat r}}\wedge_{*}
{\widetilde {\hat V}}{}^{d} -{\widetilde {\hat
V}}{}^{d}\wedge_{*}{\widetilde {\hat r}}\Bigr) +4
\eta_{dc}\Bigr(\hat V^{d}\wedge_{*} \hat r +\hat r \wedge_{*}\hat
V^{d}\Bigr)=0.
\label{(10)}
\end{eqnarray}
To second order in $\theta$, the
field equation for ${\widetilde
{\hat V}}{}^a$ is identically satisfied, while the one for
$\hat V^a$, given in (\ref{(9)}), reads as
\begin{eqnarray} &\;&
2\Bigr[-\Bigr(\hat V^{d0}\wedge_{*II}\hat R^{ab0}+\hat
V^{d2}\we\hat R^{ab0}+\hat V^{d0}\we\hat R^{ab2}\Bigr)
\varepsilon_{abcd}+{\rm i}(\eta_{bc}\eta_{ad}-\eta_{ac}\eta_{bd}) \hat
R^{ab0}\wedge_{*I}{\widetilde {\hat V}}{}^{d}
\nonumber \\
& &~~+ 4{\rm i} \eta_{dc}{\widetilde {\hat r}}
\wedge_{*I}\hat V^{d}  +4
\eta_{dc}{\widetilde {\hat V}}{}^{d}\we{\hat r} \Bigr]=0,
\label{(11)}
\end{eqnarray}

\section{Classical tetrads which solve the vacuum Einstein equations}

Since in Ref. \cite{asch2011} some of us had found that only half of the
non-commutative field equations are satisfied to first order in
$\theta$, when the classical tetrad describes the Schwarzschild
solution in the standard coordinates, we have tried to check the result
both in other coordinates for Schwarzschild, and for other solutions
of the vacuum Einstein equations.

\subsection{Classical tetrad in Kruskal--Szekeres coordinates}

The Kruskal--Szekeres coordinates are very convenients because
they make it evident that $r=0$ is only a coordinate singularity
of the Schwarzschild geometry. Following \cite{krusc},
we write $(\sigma,\tau,\theta,\phi)$ for the Kruskal--Szekeres
coordinates, with classical tetrad having components
\begin{eqnarray}
V^{(\sigma)} =&&\frac{\tau \,{\sqrt{A(r(\sigma ,\tau ))}}}
{{\sqrt{2}}\,\alpha }d\sigma
+\frac{\alpha \,{\sqrt{A(r(\sigma ,
\tau ))}}}{{\sqrt{2}}\,\tau }d\tau,
\nonumber \\
V^{(\tau)}=&&\frac{\tau \,{\sqrt{A(r(\sigma ,\tau))}}}
{{\sqrt{2}}\,\alpha }d\sigma
-\frac{\alpha \,{\sqrt{A(r(\sigma ,\tau ))}}}{{\sqrt{2}}\,\tau }d\tau,
\nonumber \\
V^{(\theta)}=&&r(\sigma,\tau)d\theta,
\nonumber \\
V^{(\tau)} =&&r(\sigma,\tau)\sin{\theta}d\phi,
\end{eqnarray}
where $A(r(\sigma ,\tau
))=\frac{4\lambda^3}{r}\exp^{-r/\lambda}
=\frac{-2\lambda^3}{r}(\frac{r}{\lambda}-1)\frac{1}{\sigma\tau}$,
and $r$ is defined implicitly by
$-\sigma\tau=(\frac{r}{\lambda}-1)\exp^{r/\lambda}$.
Hence we find
\begin{eqnarray}
&&r^{(1,0)}=\frac{-2\,{\lambda }^2\,\tau }
{{\rm e}^{\frac{r(\sigma ,\tau )}{\lambda }}\,r(\sigma ,\tau )},
\nonumber \\
&&r^{(0,1)}= \frac{-2\,{\lambda }^2\,\sigma }
{{\rm e}^{\frac{r(\sigma ,\tau )}{\lambda }}\,r(\sigma ,\tau)},
\nonumber \\
&& r^{(1,1)}=\frac{-2\,{\lambda }^2}
{{\rm e}^{\frac{r(\sigma ,\tau )}{\lambda }}\,
\left( \lambda  + r(\sigma ,\tau ) \right)},
\end{eqnarray}
and the resulting spin-connection reads as \cite{krusc}
\begin{eqnarray}
&&\omega^{01}_{1}=-\frac{1}{2}\left(1-\frac{\lambda^2}{r^2}\right),
\nonumber \\
&&\omega^{01}_{2}=-\frac{1}{2}\left(1+\frac{\lambda^2}{r^2}\right),
\nonumber \\
&&\omega^{23}_{3}=-\cos\theta,
\nonumber \\
&&\omega^{02}_{2}=-\sqrt{\frac{A}{2}}\frac{1}{2\lambda}
\left(\frac{\sigma\tau}{\alpha}+\alpha\right),
\nonumber \\
&&\omega^{03}_{3}=-\sqrt{\frac{A}{2}}\frac{\sin{\theta}}
{2\lambda}\left(\frac{\sigma\tau}{\alpha}+\alpha\right),
\nonumber \\
&&\omega^{12}_{2}=-\sqrt{\frac{A}{2}}\frac{1}{2\lambda}
\left(\frac{\sigma\tau}{\alpha}-\alpha\right),
\nonumber \\
&&\omega^{13}_{3}=-\sqrt{\frac{A}{2}}\frac{\sin{\theta}}{2\lambda}
\left(\frac{\sigma\tau}{\alpha}-\alpha\right),
\end{eqnarray}
where the labels $(1,2,3,4)$ refer to the coordinates
$(\sigma,\tau,\theta,\phi)$, respectively.

If the only non-vanishing component of non-commutativity is
$\theta^{23}$, the non-commutative field equations reduce,
to first order in $\theta^{23}$, to the form \cite{asch2011}
\begin{equation}
K_{c123} dx^{1} \wedge dx^{2} \wedge dx^{3}
+K_{c124} dx^{1} \wedge dx^{2} \wedge dx^{4}
+K_{c134} dx^{1} \wedge dx^{3} \wedge dx^{4}
+K_{c234} dx^{2} \wedge dx^{3} \wedge dx^{4}.
\end{equation}
For the first-order
field equations to hold, this should vanish identically,
which is not the case because, on the contrary,
\begin{eqnarray}
K_{c123} = &&\delta_{c1}\frac{1}{{\rm e}^{\frac{3\,r(\sigma ,\tau
)}{\lambda }}\,
\alpha \,\tau \,{r(\sigma ,\tau )}^5\,
{\left( \lambda  + r(\sigma ,\tau ) \right) }^2}
\nonumber \\
\times && 4\,{\sqrt{2}}\,\lambda \,
{\sqrt{\frac{{\lambda }^3}
{{\rm e}^{\frac{r(\sigma ,\tau )}{\lambda }}\,r(\sigma ,\tau )}}}\,
\left({\rm e}^{\frac{2\,r(\sigma ,\tau )}{\lambda }}\,
{r(\sigma ,\tau )}^5\,
\left( 3\,\lambda \,
\left({\alpha }^2 + \sigma \,\tau  \right)  +
\left(2\,{\alpha }^2 + 3\,\sigma \,\tau  \right) \,
r(\sigma ,\tau ) \right)\right.
\nonumber \\ &&+
{\rm e}^{\frac{r(\sigma ,\tau )}{\lambda }}\,\lambda \,\sigma \,
\tau \,{r(\sigma ,\tau )}^2\,
{\left( \lambda  + r(\sigma ,\tau ) \right) }^2\,
\left(-8\,{\alpha }^2\,\lambda  +
4\,\lambda \,\sigma \,\tau  +
\left(-3\,{\alpha }^2 + 5\,\sigma \,\tau  \right) \,
r(\sigma ,\tau ) \right)
\nonumber \\
&&+4\,{\lambda }^2\,{\sigma }^2\,{\tau }^2\,
{\Bigr(\lambda  + r(\sigma ,\tau ) \Bigr) }^2\,
\Bigr({\lambda }^2\,
\Bigr(-5\,{\alpha }^2 + 3\,\sigma \,\tau  \Bigr)
\nonumber \\
&-& 2\,\left(2\,{\alpha }^2 - \sigma \,\tau  \right) \,
r(\sigma ,\tau )\,
(2\,\lambda  + r(\sigma ,\tau )))\Bigr),
\nonumber \\
K_{c124} = &&\delta_{c4}{{\rm e}^
{-\frac{3\,r(\sigma ,\tau )}{\lambda }}} 8\,\lambda \,
\cos (\theta )\,
\nonumber \\
&&\left(\frac{2\,{\lambda }^4\,{\sigma }^2\,\tau \,
\left( \lambda  + r(\sigma ,\tau ) \right) }{{r(\sigma ,
\tau )}^5} + \frac{{\rm e}^
{\frac{2\,r(\sigma ,\tau )}{\lambda }}\,
\left(3\,\lambda  + r(\sigma ,\tau ) \right) }{\tau \,
\left( \lambda  + r(\sigma ,\tau ) \right) } \right.
\nonumber \\
&&+\left. \frac{{\rm e}^{\frac{r(\sigma ,\tau )}
{\lambda }}\,\lambda \,
\sigma \,\left( 2\,{\lambda }^2 +
r(\sigma ,\tau )\,
\left( \lambda  + r(\sigma ,\tau ) \right)  \right) }
{{r(\sigma ,\tau )}^3} \right),
\nonumber \\
K_{c134}=&&\delta_{c4}\frac{\lambda }{{\rm e}^
{\frac{3\,r(\sigma ,\tau )}{\lambda }}\,
{r(\sigma ,\tau )}^4\,{\left( \lambda  + r(\sigma ,\tau )
\right) }^2}
\nonumber \\
&& 4\left( -\left({\rm e}^
{\frac{2\,r(\sigma ,\tau )}{\lambda }}\,
{r(\sigma ,\tau )}^6 \right)  -
2\,{\rm e}^{\frac{r(\sigma ,\tau )}{\lambda }}\,\lambda \,
\sigma \,\tau \,{r(\sigma ,\tau )}^2\,
\left( \lambda  + r(\sigma ,\tau ) \right) \,
{\left( 2\,\lambda  + r(\sigma ,\tau ) \right) }^2 \right.
\nonumber \\
&&-\left.4\,{\lambda }^2\,{\sigma }^2\,{\tau }^2\,
{\left( \lambda  + r(\sigma ,\tau ) \right) }^2\,
\left(5\,{\lambda }^2 +
4\,r(\sigma ,\tau )\,
\left( 2\,\lambda  + r(\sigma ,\tau ) \right)  \right)
\right) \,\sin (\theta),
\nonumber \\
K_{c234} = &&\delta_{c4}\frac{1}{{\rm e}^
{\frac{3\,r(\sigma ,\tau )}{\lambda }}\,
{r(\sigma ,\tau )}^4\,\left( \lambda  + r(\sigma ,\tau )\right)}
\nonumber \\
&&8\,{\lambda }^2\,{\sigma }^2\,
\left(-\left({\rm e}^{\frac{r(\sigma ,\tau )}{\lambda }}\,
{r(\sigma ,\tau )}^2\,
\left( 2\,{\lambda }^2 + 3\,\lambda \,r(\sigma ,\tau ) +
2\,{r(\sigma ,\tau )}^2 \right)  \right) \right.
\nonumber \\
&&-\left.  2\,\lambda \,\sigma \,\tau \,
\left(\lambda  + r(\sigma ,\tau ) \right) \,
\left( 3\,{\lambda }^2 +2\,r(\sigma ,\tau )\,
\left( 2\,\lambda  + r(\sigma ,\tau ) \right)  \right)
\right) \,\sin (\theta).
\end{eqnarray}

\subsection{The Plebanski case}

We here consider the class of vacuum space-times with two commuting
Killing vector fields ${\partial \over \partial \sigma}$ and
${\partial \over \partial \tau}$ first found by Plebanski,
whose tetrad \cite{stephani} reads as
\begin{eqnarray}
V^{(\sigma)} =&&-\sqrt{C(q,p)}{\rm d}\sigma,
\nonumber \\
V^{(q)}=&&\frac{{\left( -p + q \right) }^
{\frac{1 - {\left( \beta  - \gamma  \right) }^2}{2}}\,
{\left( p + q \right) }^
{\frac{1 - {\left( \beta  + \gamma  \right) }^2}{2}}\,
{\sqrt{A(q,p)}}}{{\sqrt{-1 + q^2}}}{\rm d}q,
\nonumber \\
V^{(p)}=&&\frac{{\left( -p + q \right) }^
{\frac{1 - {\left( \beta  - \gamma  \right) }^2}{2}}\,
{\left( p + q \right) }^
{\frac{1 - {\left( \beta  + \gamma  \right) }^2}{2}}\,
{\sqrt{A(q,p)}}}{{\sqrt{1 - p^2}}}{\rm d}p,
\nonumber \\
V^{(\tau)} =&&-\sqrt{B(q,p)}{\rm d}\tau,
\end{eqnarray}
where
\begin{eqnarray}
A(q,p) & \equiv & m^2 (q+1)^{\alpha^2 -1/4}(q-1)^{\beta^2
-1/4}(p+1)^{\gamma^2 -1/4} (1-p)^{\delta^2 -1/4},
\nonumber \\
B(q,p) &\equiv & m^2 (q+1)^{\alpha +1/2}(q-1)^{\beta+1/2}(p+1)^{\gamma
+1/2} (1-p)^{\delta +1/2}
\nonumber \\
C(q,p) &\equiv & m^2 (q+1)^{1/2-\alpha
}(q-1)^{1/2-\beta}(p+1)^{1/2-\gamma } (1-p)^{1/2-\delta}
\nonumber \\
\alpha &=& a+b, \,\, \beta = a-b,\,\, \gamma \
= a+c, \,\, \delta =a-c,
\end{eqnarray}
where $a,b,c,m$ are real constants, and the coordinates $(p,q)$ lie
in the intervals $-1<p<1$, $q<\infty$, respectively.

The components of non-commutative field equations take the form
\begin{eqnarray}
K_{c123} =&&\delta_{c1}F(p,q),
\nonumber \\
K_{c124} =&& 0,
\nonumber \\
K_{c134} = &&0,
\nonumber \\
K_{c234} =&&\delta_{c4}G(p,q),
\end{eqnarray}
where
\begin{eqnarray}
&&F(p,q) \equiv \frac{1}{2\,{\sqrt{-1 + {p}^2}}\,
    {\left( p  - q  \right) }^2\,
    {\left( p  + q  \right) }^2\,{\sqrt{-1 + {q }^2}}\,
    {A(p ,q )}^2}\nonumber \\
     \times &&\left( \frac{-\left( {\sqrt{-1 + {q }^2}}\,
           \left( 2\,\left( -2\,\beta \,\gamma \,p  +
                {\beta }^2\,q  +
                \left( -1 + {\gamma }^2 \right) \,q  \right) \,
              A(p,q) +
             \left( {p}^2 - {q}^2 \right) \,
              A^{(0,1)}(p ,q ) \right) \,
           C^{(0,1)}(p ,q ) \right) }{4\,
         {\sqrt{-1 + {p }^2}}\,
         \left( p  - q  \right) \,
         \left( p  + q  \right) \,A(p,q)\,
         {\sqrt{C(p ,q )}}} \right.\nonumber \\ &&
      +\left.\frac{{\sqrt{-1 + {p }^2}}\,
         \left( -2\,\left( \left( -1 + {\beta }^2 +
                 {\gamma }^2 \right) \,p  -
              2\,\beta \,\gamma \,q  \right) \,A(p ,q )
            + \left( {p }^2 - {q }^2 \right) \,
            A^{(1,0)}(p ,q ) \right) \,
         C^{(1,0)}(p ,q )}{4\,
         \left( -p  + q  \right) \,
         \left( p  + q  \right) \,
         {\sqrt{-1 + {q }^2}}\,A(p ,q )\,
         {\sqrt{C(p ,q )}}} \right) \,
    \nonumber \\\times && \left( 4\,\left( 2\,\beta \,
\gamma \,{p }^3\,q  +
         \left( -1 + {\beta }^2 + {\gamma }^2 \right) \,
          {q }^2 + 2\,\beta \,\gamma \,p \,q \,
          \left( -2 + {q }^2 \right)\right.\right.  \nonumber \\ && -
         \left.\left( -1 + {\beta }^2 + {\gamma }^2 \right) \,
          {p}^2\,\left( -1 + 2\,{q}^2 \right)  \right){A(p,q
          )}^2
          \nonumber \\ &&+
    \left.  {\left( {p}^2 - {q}^2 \right) }^2\,
       \left( \left( -1 + {q}^2 \right) \,
          {A^{(0,1)}(p,q)}^2 +
         \left( -1 + {p}^2 \right) \,
          {A^{(1,0)}(p,q)}^2 \right)\right.
          \nonumber \\ &&
          -
      {\left( {p}^2 - {q}^2 \right) }^2\,
       A(p,q)\,\left(q \,A^{(0,1)}(p,q) +
         \left( -1 + {q}^2 \right) \,
          A^{(0,2)}(p ,q ) +
         p \,A^{(1,0)}(p,q)\right.\nonumber \\ && -
        \left.\left. A^{(2,0)}(p,q) +
         {p}^2\,A^{(2,0)}(p,q) \right)  \right),
    \end{eqnarray}
while $G$ is given by too lengthy an expression for a regular
paper.

\subsection{Vacuum solutions of class AIII}

The degenerate static vacuum solutions of class AIII
\cite{stephani} can be described by using the coordinates
$(t,r,z,\phi)$ to which no Killing vectors can be associated.
The corresponding classical tetrad has components \cite{stephani}
\begin{eqnarray}
V^{(t)} =&&-\frac{1}{\sqrt{z}}{\rm d}t,
\nonumber \\
V^{(z)}=&&\sqrt{z}\; {\rm d}z,
\nonumber \\
V^{(r)}=&&z\; {\rm d}r,
\nonumber \\
V^{(\phi)} =&& rz {\rm d}\phi.
\end{eqnarray}
To first order in non-commutativity, the components of field equations
turn out to be
\begin{eqnarray}
K_{c123} =&&\delta_{c1}\frac{-5}{z^{\frac{7}{2}}},
\nonumber \\
K_{c124} =&& 0,
\nonumber \\
K_{c134} = &&0,
\nonumber \\
K_{c234} =&&\delta_{c4}\frac{-2\,r}{z^2}.
\end{eqnarray}
At this stage, one might think that the problem lies in the attempt
of finding solutions of the non-commutative field equations via the
Seiberg--Witten map. For this purpose, we show in the appendix what
happens if the Seiberg--Witten map is not used for finding solutions
of non-commutative field equations.

\section{The simpler case of vacuum Maxwell theory}

Since in \cite{asch2011} it had been found that inserting the
Seiberg--Witten map in the non-commutative field equations is
inequivalent to inserting it directly in the action for pure gravity,
we here consider the simpler case of vacuum Maxwell theory, for which
the non-commutative potential, gauge parameter and field strength
read as
\begin{equation}
{\widehat A}_{\mu}=A_{\mu}+{\widetilde A}_{\mu}
+{\rm O}(\theta^{2}),
\end{equation}
\begin{equation}
{\widehat \Lambda}=\Lambda+{\widetilde \Lambda}
+{\rm O}(\theta^{2}),
\end{equation}
\begin{equation}
{\widehat F}_{\mu \nu}=\partial_{\mu}{\widehat A}_{\nu}
-\partial_{\nu}{\widehat A}_{\mu}
-{\rm i}[{\widehat A}_{\mu},{\widehat A}_{\nu}]_{\star}
=F_{\mu \nu}^{(0)}+{\widetilde F}_{\mu \nu}
+{\rm O}(\theta^{2}),
\end{equation}
where
\begin{equation}
{\widetilde F}_{\mu \nu}=\partial_{\mu}{\widetilde A}_{\nu}
-\partial_{\nu}{\widetilde A}_{\mu}
+\theta^{\rho \sigma}\partial_{\rho}A_{\mu}
\partial_{\sigma}A_{\nu}.
\end{equation}
The equation defining the Seiberg--Witten map
\begin{equation}
{\widehat \delta}_{\widehat \Lambda}{\widehat A}_{\mu}(A)
={\widehat A}_{\mu}(A+\delta_{\Lambda}A)-{\widehat A}_{\mu}(A)
\end{equation}
is solved, to first order in $\theta^{\rho \sigma}$, by
\begin{equation}
{\widehat A}_{\mu}=-{\theta^{\rho \sigma}\over 2}
\Bigr(A_{\rho}\partial_{\sigma}A_{\mu}
+A_{\rho}F_{\sigma \mu}\Bigr),
\end{equation}
\begin{equation}
{\widehat \Lambda}={\theta^{\rho \sigma}\over 2}
(\partial_{\rho}\Lambda)A_{\sigma}.
\end{equation}
Variation of the action functional to first order in $\theta$
yields now equations in agreement with the insertion of the
Seiberg--Witten map in the non-commutative field equations
\begin{equation}
{\widehat D}^{\mu}{\widehat F}_{\mu \nu}=0.
\end{equation}
In both cases, we find eventually
\begin{equation}
\theta^{\rho \sigma}u_{\rho \sigma \nu}=0,
\end{equation}
where
\begin{eqnarray}
u_{\rho \sigma \nu}&=& -(\Box A_{\rho})(\partial_{\sigma}A_{\nu})
+{1\over 2}\biggr[(\Box A_{\rho})(\partial_{\nu}A_{\sigma})
-(\Box A_{\sigma})(\partial_{\nu}A_{\rho})\biggr]
\nonumber \\
&+&{1\over 2}A_{\rho}\biggr[\partial_{\nu}(\Box A_{\sigma})
-\partial_{\sigma}(\Box A_{\nu})\biggr]
\nonumber \\
&+& (\partial_{\mu}A_{\rho})\biggr[{1\over 2}
\partial^{\mu}\partial_{\nu}A_{\sigma}
-\partial^{\mu}\partial_{\sigma}A_{\nu}\biggr]
\nonumber \\
&+& (\partial^{\mu}A_{\rho})\biggr[\partial_{\nu \sigma}^{2}A_{\mu}
-\partial_{\mu \sigma}^{2}A_{\nu}\biggr]
\nonumber \\
&+& {1\over 2}A_{\rho}\biggr[2 \partial^{\mu}\partial_{\nu \sigma}^{2}
A_{\mu}+\partial^{\mu}(\partial_{\mu \sigma}^{2}A_{\nu}
-\partial_{\mu \nu}^{2}A_{\sigma})\biggr]
\nonumber \\
&+& (\partial^{\mu}\partial_{\nu}A_{\rho})
\biggr[\partial_{\sigma}A_{\mu}
-{1\over 2}\partial_{\mu}A_{\sigma}\biggr]
\nonumber \\
&+& {1\over 2}(\partial_{\nu}A_{\rho})
\biggr[2 \partial^{\mu}\partial_{\sigma}A_{\mu}
-\Box A_{\sigma}\biggr]
\nonumber \\
&+& (\partial_{\rho}A^{\mu})(\partial_{\mu \sigma}^{2}A_{\nu}
-\partial_{\nu \sigma}^{2}A_{\mu})
\nonumber \\
&+& \biggr[(\partial_{\rho}A_{\mu})(\partial^{\mu}\partial_{\sigma}
A_{\nu})+(\partial_{\sigma}A_{\nu})
(\partial^{\mu}\partial_{\rho}A_{\mu})\biggr].
\end{eqnarray}
This is a sort of {\it off-shell formula}, because we have not assumed
that the classical background solves the vacuum Maxwell equations.
For example, in the Lorenz gauge $\partial^{\nu}A_{\nu}=0$,
$A_{\mu}$ obeys the wave equation $\Box A_{\mu}=0$, and
$u_{\rho \sigma \nu}$ takes the {\it on-shell form}
\begin{eqnarray}
{\tilde u}_{\rho \sigma \nu}&=&
{1\over 2}(\partial_{\mu}A_{\rho})
\Bigr(\partial^{\mu}\partial_{\nu}A_{\sigma}
-2\partial^{\mu}\partial_{\sigma}A_{\nu}\Bigr)
\nonumber \\
&+& (\partial^{\mu}A_{\rho})(\partial_{\nu \sigma}^{2}A_{\mu}
-\partial_{\mu \sigma}^{2}A_{\nu})
\nonumber \\
&+&{1\over 2}A_{\rho}\partial^{\mu}(\partial_{\mu \sigma}^{2}
A_{\nu}-\partial_{\mu \nu}^{2}A_{\sigma})
\nonumber \\
&+& {1\over 2}(\partial^{\mu}\partial_{\nu}A_{\rho})
\Bigr(2 \partial_{\sigma}A_{\mu}-\partial_{\mu}A_{\sigma}\Bigr)
\nonumber \\
&+& (\partial_{\rho}A^{\mu})\Bigr(\partial_{\mu \sigma}^{2}
A_{\nu}-\partial_{\nu \sigma}^{2}A_{\mu}\Bigr)
\nonumber \\
&+& (\partial_{\rho}A_{\mu})\Bigr(\partial^{\mu}
\partial_{\sigma}A_{\nu}).
\end{eqnarray}
Eventually, if $\theta^{\rho \sigma}$ reads as \cite{asch2011}
\begin{equation}
\theta^{\rho \sigma}=\theta \Bigr(\delta^{\rho 2} \delta^{\sigma 3}
-\delta^{\rho 3} \delta^{\sigma 2}\Bigr),
\end{equation}
we have to check whether $\theta^{\rho \sigma} {\tilde u}_{\rho
\sigma \nu}$ vanishes when $\Box A_{\mu}=0$. In Cartesian
coordinates, on considering the simple plane-wave solution to
Maxwell's equations in Lorenz gauge, i.e.
\begin{equation}
A_\mu(t,x,y,z)={\rm e}^{{\rm i}(\omega t - k\cdot x)}U_{\mu},
\end{equation}
the components of $u_\nu=\theta^{\rho \sigma}
{\tilde u}_{\rho \sigma \nu}$ turn out to be
\begin{eqnarray}
u_1=&& -\frac{{\rm i}}{2}\,{\rm e}^
   {2\,{\rm i}\left(  \omega t -
     \,
      \left( x\,{{\kappa }_1} + y\,{{\kappa }_2} +
        z\,{{\kappa }_3} \right) \right)}\,
  \left( 4\,{{\kappa }_1}\,
     \left( \omega \,{U_4} + {U_2}\,{{\kappa }_2} +
       {U_3}\,{{\kappa }_3} \right)  +
    {U_1}\,\left( 3\,{\omega }^2 + {{{\kappa }_1}}^2 -
       3\,{{{\kappa }_2}}^2 - 3\,{{{\kappa }_3}}^2 \right)
    \right) \,\nonumber \\ &&\times\left( -\left( {U_3}\,{{\kappa }_1}\,
       {{\theta }_{13}} \right)  +
    {U_4}\,{{\kappa }_1}\,{{\theta }_{14}} +
    {U_1}\,\left( {{\kappa }_2}\,{{\theta }_{12}} +
       {{\kappa }_3}\,{{\theta }_{13}} +
       \omega \,{{\theta }_{14}} \right)\right.  \nonumber \\ &&-
   \left. {U_3}\,{{\kappa }_2}\,{{\theta }_{23}} +
    {U_4}\,{{\kappa }_2}\,{{\theta }_{24}} +
    {U_2}\,\left( -\left( {{\kappa }_1}\,{{\theta }_{12}} \right)
           + {{\kappa }_3}\,{{\theta }_{23}} +
       \omega \,{{\theta }_{24}} \right)  +
    \omega \,{U_3}\,{{\theta }_{34}} +
    {U_4}\,{{\kappa }_3}\,{{\theta }_{34}} \right),
\nonumber \\
u_2=&& -\frac{{\rm i}}{2}\,{\rm e}^
   {2\,{\rm i}\left(  \omega t -
     \,
      \left( x\,{{\kappa }_1} + y\,{{\kappa }_2} +
        z\,{{\kappa }_3} \right) \right)}\,
  \left( 4\,{{\kappa }_2}\,
     \left( \omega \,{U_4} + {U_1}\,{{\kappa }_1} +
       {U_3}\,{{\kappa }_3} \right)  +
    {U_2}\,\left( 3\,{\omega }^2 - 3\,{{{\kappa }_1}}^2 +
       {{{\kappa }_2}}^2 - 3\,{{{\kappa }_3}}^2 \right)  \right)
   \nonumber \\ && \times\left( -\left( {U_3}\,{{\kappa }_1}\,{{\theta }_{13}} \right)
        + {U_4}\,{{\kappa }_1}\,{{\theta }_{14}} +
    {U_1}\,\left( {{\kappa }_2}\,{{\theta }_{12}} +
       {{\kappa }_3}\,{{\theta }_{13}} +
       \omega \,{{\theta }_{14}} \right) \right. \nonumber \\ &&-
   \left. {U_3}\,{{\kappa }_2}\,{{\theta }_{23}} +
    {U_4}\,{{\kappa }_2}\,{{\theta }_{24}} +
    {U_2}\,\left( -\left( {{\kappa }_1}\,{{\theta }_{12}} \right)
           + {{\kappa }_3}\,{{\theta }_{23}} +
       \omega \,{{\theta }_{24}} \right)  +
    \omega \,{U_3}\,{{\theta }_{34}} +
    {U_4}\,{{\kappa }_3}\,{{\theta }_{34}} \right),
\nonumber \\
u_3=&& -\frac{{\rm i}}{2}\,{\rm e}^
   {2\,{\rm i}\left(  \omega t -
     \,
      \left( x\,{{\kappa }_1} + y\,{{\kappa }_2} +
        z\,{{\kappa }_3} \right) \right)}\,
  \left( 4\,\left( \omega \,{U_4} + {U_1}\,{{\kappa }_1} +
       {U_2}\,{{\kappa }_2} \right) \,{{\kappa }_3} +
    {U_3}\,\left( 3\,{\omega }^2 - 3\,{{{\kappa }_1}}^2 -
       3\,{{{\kappa }_2}}^2 + {{{\kappa }_3}}^2 \right)  \right) \
   ,\nonumber \\&&\times\left( -\left( {U_3}\,{{\kappa }_1}\,{{\theta }_{13}} \right)
        + {U_4}\,{{\kappa }_1}\,{{\theta }_{14}} +
    {U_1}\,\left( {{\kappa }_2}\,{{\theta }_{12}} +
       {{\kappa }_3}\,{{\theta }_{13}} +
       \omega \,{{\theta }_{14}} \right) \right.\nonumber \\ && \left. -
    {U_3}\,{{\kappa }_2}\,{{\theta }_{23}} +
    {U_4}\,{{\kappa }_2}\,{{\theta }_{24}} +
    {U_2}\,\left( -\left( {{\kappa }_1}\,{{\theta }_{12}} \right)
           + {{\kappa }_3}\,{{\theta }_{23}} +
       \omega \,{{\theta }_{24}} \right)  +
    \omega \,{U_3}\,{{\theta }_{34}} +
    {U_4}\,{{\kappa }_3}\,{{\theta }_{34}} \right),
\nonumber \\
      u_4= && -\frac{{\rm i}}{2}\,{\rm e}^
   {2\,{\rm i}\left(  \omega t -
     \,
      \left( x\,{{\kappa }_1} + y\,{{\kappa }_2} +
        z\,{{\kappa }_3} \right) \right)}\,
  \left( 4\,\omega \,\left( {U_1}\,{{\kappa }_1} +
       {U_2}\,{{\kappa }_2} + {U_3}\,{{\kappa }_3} \right)  +
    {U_4}\,\left( {\omega }^2 + 3\,{{{\kappa }_1}}^2 +
       3\,{{{\kappa }_2}}^2 + 3\,{{{\kappa }_3}}^2 \right)
    \right) \,\nonumber \\ &&\left( -\left( {U_3}\,{{\kappa }_1}\,
       {{\theta }_{13}} \right)  +
    {U_4}\,{{\kappa }_1}\,{{\theta }_{14}} +
    {U_1}\,\left( {{\kappa }_2}\,{{\theta }_{12}} +
       {{\kappa }_3}\,{{\theta }_{13}} +
       \omega \,{{\theta }_{14}} \right)\right.  \nonumber \\ &&\left.-
    {U_3}\,{{\kappa }_2}\,{{\theta }_{23}} +
    {U_4}\,{{\kappa }_2}\,{{\theta }_{24}} +
    {U_2}\,\left( -\left( {{\kappa }_1}\,{{\theta }_{12}} \right)
           + {{\kappa }_3}\,{{\theta }_{23}} +
       \omega \,{{\theta }_{24}} \right)  +
    \omega \,{U_3}\,{{\theta }_{34}} +
    {U_4}\,{{\kappa }_3}\,{{\theta }_{34}} \right).
    \label{1}
\end{eqnarray}
In the case $\theta_{23}\not = 0$, we find
\begin{eqnarray}
u_1=&& \frac{{\rm i}}{2}\,{\rm e}^
   {2\,{\rm i}\left(  \omega t -
     \,
      \left( x\,{{\kappa }_1} + y\,{{\kappa }_2} +
        z\,{{\kappa }_3} \right) \right)}\,{{\theta }_{23}}\,
  \left( -{U_3}\,{{\kappa }_2} + {U_2}\,{{\kappa }_3} \right) \,
  \left( 4\,{{\kappa }_1}\,
     \left( \omega \,{U_4} + {U_2}\,{{\kappa }_2} +
       {U_3}\,{{\kappa }_3} \right)  \right.\nonumber \\ &&\left.+
    {U_1}\,\left( 3\,{\omega }^2 + {{{\kappa }_1}}^2 -
       3\,{{{\kappa }_2}}^2 - 3\,{{{\kappa }_3}}^2 \right)
    \right),
\nonumber \\
u_2=&& \frac{{\rm i}}{2}\,{\rm e}^
   {2\,{\rm i}\left(  \omega t -
     \,
      \left( x\,{{\kappa }_1} + y\,{{\kappa }_2} +
        z\,{{\kappa }_3} \right) \right)}\,{{\theta }_{23}}\,
  \left( -\left( {U_3}\,{{\kappa }_2} \right)  +
    {U_2}\,{{\kappa }_3} \right) \,
  \left( 4\,{{\kappa }_2}\,
     \left( \omega \,{U_4} + {U_1}\,{{\kappa }_1} +
       {U_3}\,{{\kappa }_3} \right) \right.\nonumber \\ &&\left. +
    {U_2}\,\left( 3\,{\omega }^2 - 3\,{{{\kappa }_1}}^2 +
       {{{\kappa }_2}}^2 - 3\,{{{\kappa }_3}}^2 \right)  \right),
\nonumber \\
u_3=&& \frac{{\rm i}}{2}\,{\rm e}^
   {2\,{\rm i}\left(  \omega t -
     \,
      \left( x\,{{\kappa }_1} + y\,{{\kappa }_2} +
        z\,{{\kappa }_3} \right) \right)}\,{{\theta }_{23}}\,
  \left(- {U_3}\,{{\kappa }_2} + {U_2}\,{{\kappa }_3} \right) \,
  \left( 4\,\left( \omega \,{U_4} + {U_1}\,{{\kappa }_1} +
       {U_2}\,{{\kappa }_2} \right) \,{{\kappa }_3} \right.
\nonumber \\ 
&&\left.  +
    {U_3}\,\left( 3\,{\omega }^2 - 3\,{{{\kappa }_1}}^2 -
       3\,{{{\kappa }_2}}^2 + {{{\kappa }_3}}^2 \right)  \right),
\nonumber \\
u_4 = && \frac{{\rm i}}{2}\,{\rm e}^
   {2\,{\rm i}\left(  \omega t -
     \,
      \left( x\,{{\kappa }_1} + y\,{{\kappa }_2} +
        z\,{{\kappa }_3} \right) \right)}\,{{\theta }_{23}}\,
  \left( {U_3}\,{{\kappa }_2} - {U_2}\,{{\kappa }_3} \right) \,
  \left( 4\,\omega \,\left( {U_1}\,{{\kappa }_1} +
       {U_2}\,{{\kappa }_2} + {U_3}\,{{\kappa }_3} \right) \right.
\nonumber \\ 
&&\left. +
    {U_4}\,\left( {\omega }^2 + 3\,{{{\kappa }_1}}^2 +
       3\,{{{\kappa }_2}}^2 + 3\,{{{\kappa }_3}}^2 \right)
    \right),
\end{eqnarray}
and hence $u_\nu=\theta^{\rho \sigma}
{\tilde u}_{\rho \sigma \nu}=0$ is satisfied when
\begin{equation}
\frac{U_3}{U_2}= \frac{{\kappa  }_3}{{\kappa }_2}.
\end{equation}
For all $\theta_{ij}\not = 0$, $u_\nu=\theta^{\rho \sigma} {\tilde
u}_{\rho \sigma \nu}=0$ (see Eq. (\ref{1})) is satisfied when
\begin{eqnarray}
&&U_i= \kappa_i, \,\, i=1,2,3, 
\nonumber \\
&&U_4= -\omega .
\end{eqnarray}
However, this is too particular a solution to first order, and its
physical meaning, if any, is unclear to us.

In polar coordinates, on considering the simple spherical-wave
solution to Maxwell's equations in Lorenz gauge, i.e.
\begin{equation}
A_\mu(t,r)=\frac{U_{\mu}}{r}{\rm e}^{{\rm i}(\omega t - |\kappa|
r)},
\end{equation}
in the case $\theta_{23}\not = 0$, $u_\nu=\theta^{\rho \sigma}
{\tilde u}_{\rho \sigma \nu}=0$ is satisfied.

\section{Concluding remarks}

Our paper has investigated relevant topics in non-commutative gravity \cite{MS06}-\cite{RDS12} and noncommutative gauge theories \cite{BK05}-\cite{LY04}. For pure gravity, the non-commutative field equations first found
in Ref. \cite{asch2011} cannot be both solved by inserting the
Seiberg--Witten map, nor do we succeed in finding solutions which do
not exploit such a map, as is shown in detail in the appendix.
This is now tested for three backgrounds
solving the vacuum Einstein equations. The second-order Seiberg--Witten
map for pure gravity has also been evaluated explicitly in our paper
(but we acknowledge the neater geometric results first
obtained in Ref. \cite{July2012}).

If one considers instead the non-commutative version of vacuum
Maxwell theory, the stage at which the Seiberg--Witten map is
inserted does not make any difference, but no obvious solution
of the non-commutative field equations has been obtained so far
by us on this side. Moreover, the gravitational counterpart of the
off-shell versus on-shell analysis of Sec. VIII is lacking at present,
and is also a topic for further investigation in our opinion.

As far as we can see, while much progress has been made in the literature
on the Seiberg--Witten map to various orders
and its geometric structure \cite{July2012},
it remains unclear whether the non-commutative field equations
admit solutions. So far, the Seiberg--Witten map has been exploited
to establish a correspondence between non-commutative and commutative
fields, hence showing that
once the non-commutative action is expanded in
terms of the commutative fields, the resulting action is gauge-invariant
under ordinary local Lorentz transformations \cite{1207.5060, lizzi}.
But our detailed calculations show that the problem of solving
the non-commutative field equations remains open.

\appendix
\section{Studying the non-commutative field equations without the
Seiberg--Witten map}

Let us start from the non-commutative field equation written in
the form \cite{asch2011}
\begin{equation}
\Bigr[-\varepsilon_{abcd}{\widetilde V}_{\mu}^{d}R_{\nu
\lambda}^{(0)ab} +\theta^{\rho
\sigma}\Bigr(\partial_{\rho}V_{\mu}^{d}\Bigr)
\Bigr(\partial_{\sigma}R_{dc \; \nu \lambda}^{(0)}\Bigr) +4V_{c
\mu}r_{\nu \lambda}^{(1)}\Bigr] dx^{\mu} \wedge dx^{\nu} \wedge
dx^{\lambda} +{\rm O}(\theta^{2})=0, \label{(32)}
\end{equation}
where we consider ${\widetilde V}_{\mu}^{d}$ as represented by a
generic matrix
\begin{equation}
\pmatrix{{F_{1}^1}(t,r,\theta ,\phi )&
{F_{2}^1}(t,r,\theta ,\phi)&
{F^{1}_3}(t,r,\theta ,\phi )&{F^{1}_4}(t,r,\theta ,\phi )\cr
{F^{2}_1}(t,r,\theta ,\phi )& {F^{2}_2}(t,r,\theta ,\phi )&
{F^{2}_3}(t,r,\theta ,\phi )&{F^{2}_4}(t,r,\theta ,\phi )\cr
{F^{3}_1}(t,r,\theta ,\phi )& {F^{3}_2}(t,r,\theta ,\phi )&
{F^{3}_3}(t,r,\theta ,\phi )& {F^{3}_4}(t,r,\theta ,\phi )\} \cr
{F^{4}_1}(t,r,\theta ,\phi ) & {F^{4}_2}(t,r,\theta ,\phi )&
{F^{4}_3}(t,r,\theta ,\phi )& {F^{4}_4}(t,r,\theta ,\phi )\cr},
\end{equation}
and we write the components of $\omega_\mu$ as
$(\omega_1,\omega_2,\omega_3,\omega_4)$. In a Schwarzschild
background, when we require fulfillment of the torsion constraint
\begin{equation}
T_{(cl)}^{a} \wedge_{*} V^{b} \eta_{ab} \gamma_{5}
+\Bigr(T_{(cl)}^{a}\wedge {\widetilde V}^{b} -{\widetilde
T}^{a}\wedge V^{b}\Bigr)\eta_{ab} +{{\rm i}\over 2}T_{(I)}^{c} \wedge
V^{d} \varepsilon_{cdab} \gamma^{ab}=0,
\label{(16)}
\end{equation}
then its component along $\gamma^{ab}$, i.e. $G_{ab}[a, b,
\mu,\nu,\sigma]={{\rm i}\over 2}T_{(\mu\nu I)}^{c} \wedge V^{d}_\sigma
\varepsilon_{cdab}$, should vanish.
On the other hand, we find (writing for simplicity $(F_{ij}=F^i_j)$)
\begin{eqnarray}
&&G_{ab}[1, 2, 2,3,4]=4\,r^2\,\sin (\theta )\,
{{\omega }_2}(t,r,\theta ,\phi ),
\nonumber \\ &&
G_{ab}[1, 2, 1,3,4]=\frac{-\left( M\,{F_{34}}(t,r,\theta ,\phi ) -
M\,\sin (\theta )\,{F_{43}}(t,r,\theta ,\phi ) +
        8\,r^3\,\sin (\theta )\,{{\omega }_1}(t,r,\theta ,\phi )
        \right) }{2\,r} ,
\nonumber \\ &&
G_{ab}[1, 2, 1,2,4]=\frac{-\left( M\,\sin (\theta )\,
        {F_{42}}(t,r,\theta ,\phi ) \right) }{2\,r} ,
\nonumber \\ &&
G_{ab}[1, 2, 1,2,3]=\frac{M\,{F_{32}}(t,r,\theta ,\phi )}{2\,r} ,
\nonumber \\ &&
G_{ab}[1, 3, 2,3,4]=\frac{-{F_{14}}(t,r,\theta ,\phi )}{2} +
    \frac{4\,{\sqrt{1 - \frac{2\,M}{r}}}\,r^2\,\sin (\theta )\,
       {{\omega }_3}(t,r,\theta ,\phi )}{-2\,M + r} ,
\nonumber \\ &&
G_{ab}[1, 3, 1,2,4]=\frac{-\left( M\,{F_{34}}(t,r,\theta ,\phi ) +
        8\,r^3\,\sin (\theta )\,{{\omega }_1}(t,r,\theta ,\phi )
        \right) }{2\,{\sqrt{1 - \frac{2\,M}{r}}}\,r^2} ,
\nonumber \\ &&
G_{ab}[1, 3, 1,2,3]=\frac{{F_{11}}(t,r,\theta ,\phi ) +
      \frac{M\,{F_{33}}(t,r,\theta ,\phi )}
{{\sqrt{1 - \frac{2\,M}{r}}}\,r^2}}{2} ,
\nonumber \\ &&
G_{ab}[1, 4, 2,3,4]=\frac{-\left( r\,\cos (\theta )\,
         {F_{12}}(t,r,\theta ,\phi ) \right)  +
      \sin (\theta )\,{F_{13}}(t,r,\theta ,\phi ) +
      \frac{8\,{\sqrt{1 - \frac{2\,M}{r}}}\,r^2\,
         {{\omega }_4}(t,r,\theta ,\phi )}{-2\,M + r}}{2} ,
\nonumber \\ &&
G_{ab}[1, 4, 1,3,4]=\frac{r\,\cos (\theta )\,
      {F_{11}}(t,r,\theta ,\phi )}{2} ,
\nonumber \\ &&
G_{ab}[1, 4, 1,2,4]=\frac{-\left( \sin (\theta )\,
         {F_{11}}(t,r,\theta ,\phi ) \right)  -
      \frac{M\,{F_{44}}(t,r,\theta ,\phi )}
       {{\sqrt{1 - \frac{2\,M}{r}}}\,r^2}}{2} ,
\nonumber \\ &&
G_{ab}[1, 4, 1,2,3]=\frac{M\,{F_{43}}(t,r,\theta ,\phi ) -
      8\,r^3\,{{\omega }_1}(t,r,\theta ,\phi )}{2\,
      {\sqrt{1 - \frac{2\,M}{r}}}\,r^2} ,
\nonumber \\ &&
G_{ab}[2, 1, 2,3,4]=-4\,r^2\,\sin (\theta )\,
    {{\omega }_2}(t,r,\theta ,\phi ) ,
\nonumber \\ &&
G_{ab}[2, 1, 1,3,4]=\frac{M\,{F_{34}}(t,r,\theta ,\phi ) -
      M\,\sin (\theta )\,{F_{43}}(t,r,\theta ,\phi ) +
      8\,r^3\,\sin (\theta )\,{{\omega }_1}(t,r,\theta ,\phi )}
      {2\,r} ,
\nonumber \\ &&
G_{ab}[2, 1, 1,2,4]=
   \frac{M\,\sin (\theta )\,{F_{42}}(t,r,\theta ,\phi )}{2\,r} ,
\nonumber \\ &&
G_{ab}[2, 1, 1,2,3]=\frac{-\left( M\,
{F_{32}}(t,r,\theta ,\phi ) \right) }{2\,r} ,
\nonumber \\ &&
G_{ab}[2, 3, 2,3,4]=\frac{-\left( {\sqrt{1 - \frac{2\,M}{r}}}\,r\,
        \sin (\theta )\,{F_{42}}(t,r,\theta ,\phi ) \right) }{2},
\nonumber \\
&&G_{ab}[2, 3, 1,3,4]=\frac{\left( 2\,M - r \right) \,
       {F_{14}}(t,r,\theta ,\phi ) +
      {\sqrt{1 - \frac{2\,M}{r}}}\,r^2\,\sin (\theta )\,
       \left( {F_{41}}(t,r,\theta ,\phi ) +
8\,{{\omega }_3}(t,r,\theta ,\phi ) \right) }{2\,r} ,
\nonumber \\ &&
G_{ab}[2, 3, 1,2,4]=-4\,{\sqrt{1 - \frac{2\,M}{r}}}\,r\,
    \sin (\theta )\,{{\omega }_2}(t,r,\theta ,\phi ) ,
\nonumber \\ &&
G_{ab}[2, 3, 1,2,3]=\frac{\left( 1 - \frac{2\,M}{r} \right) \,
{F_{12}}(t,r,\theta ,\phi )}{2} ,
\nonumber \\ &&
G_{ab}[2, 4, 2,3,4]=\frac{-\left( r\,
        \left( \cos (\theta )\,{F_{22}}(t,r,\theta ,\phi ) -
          {\sqrt{1 - \frac{2\,M}{r}}}\,\sin (\theta )\,
           {F_{32}}(t,r,\theta ,\phi ) \right)  \right) }{2} ,
\nonumber \\ &&
G_{ab}[2, 4, 1,3,4]=\frac{1}{2\,r}\left( -2\,M + r \right) \,
       \sin (\theta )\,{F_{13}}(t,r,\theta ,\phi)
\nonumber \\ &&+
      r^2\,\left( \cos (\theta )\,{F_{21}}(t,r,\theta ,\phi ) +
         {\sqrt{1 - \frac{2\,M}{r}}}\,
          \left( -\left( \sin (\theta )\,
               {F_{31}}(t,r,\theta ,\phi ) \right)  +
            8\,{{\omega }_4}(t,r,\theta ,\phi ) \right)  \right),
\nonumber \\ &&
G_{ab}[2, 4, 1,2,4]=
   \frac{-\left( \left( 1 - \frac{2\,M}{r} \right) \,
        \sin (\theta )\,{F_{12}}(t,r,\theta ,\phi ) \right) }{2},
\nonumber \\ &&
G_{ab}[2, 4, 1,2,3]=-4\,{\sqrt{1 - \frac{2\,M}{r}}}\,r\,
    {{\omega }_2}(t,r,\theta ,\phi ) ,
\nonumber \\ &&
G_{ab}[3, 1, 2,3,4]=\frac{{F_{14}}(t,r,\theta ,\phi )}{2} -
    \frac{4\,{\sqrt{1 - \frac{2\,M}{r}}}\,r^2\,\sin (\theta )\,
       {{\omega }_3}(t,r,\theta ,\phi )}{-2\,M + r} ,
\nonumber \\ &&
G_{ab}[3, 1, 1,2,4]=\frac{M\,{F_{34}}(t,r,\theta ,\phi ) +
      8\,r^3\,\sin (\theta )\,{{\omega }_1}(t,r,\theta ,\phi )}
      {2\,{\sqrt{1 - \frac{2\,M}{r}}}\,r^2} ,
\nonumber \\ &&
G_{ab}[3, 1, 1,2,3]=\frac{-{F_{11}}(t,r,\theta ,\phi ) -
      \frac{M\,{F_{33}}(t,r,\theta ,\phi )}
       {{\sqrt{1 - \frac{2\,M}{r}}}\,r^2}}{2} ,
\nonumber \\ &&
G_{ab}[3, 2, 2,3,4]=\frac{{\sqrt{1 - \frac{2\,M}{r}}}\,r\,
      \sin (\theta )\,{F_{42}}(t,r,\theta ,\phi )}{2},
\nonumber \\ &&
G_{ab}[3, 2, 1,3,4]=\frac{\left( -2\,M + r \right) \,
       {F_{14}}(t,r,\theta ,\phi ) -
      {\sqrt{1 - \frac{2\,M}{r}}}\,r^2\,\sin (\theta )\,
       \left( {F_{41}}(t,r,\theta ,\phi ) +
         8\,{{\omega }_3}(t,r,\theta ,\phi ) \right) }{2\,r},
\nonumber \\ &&
G_{ab}[3, 2, 1,2,4]=4\,{\sqrt{1 - \frac{2\,M}{r}}}\,r\,
    \sin (\theta )\,{{\omega }_2}(t,r,\theta ,\phi ),
\nonumber \\ &&
G_{ab}[3, 2, 1,2,3]=\frac{-\left( \left( 1 - \frac{2\,M}{r} \right)
\,{F_{12}}(t,r,\theta ,\phi ) \right) }{2} ,
\nonumber \\ &&
G_{ab}[3, 4, 2,3,4]=\frac{\frac{{\sqrt{1 - \frac{2\,M}{r}}}\,r\,
         \cos (\theta )\,{F_{23}}(t,r,\theta ,\phi )}{2\,M - r} +
      \sin (\theta )\,{F_{33}}(t,r,\theta ,\phi ) +
      {F_{44}}(t,r,\theta ,\phi )}{2} ,
\nonumber \\ &&
G_{ab}[3, 4, 1,3,4]=\frac{{\sqrt{1 - \frac{2\,M}{r}}}\,
\cos (\theta )\,{F_{13}}(t,r,\theta ,\phi )}{2} ,
\nonumber \\ &&
G_{ab}[3, 4, 1,2,4]=\frac{1}{2\,{\sqrt{1
- \frac{2\,M}{r}}}\,r}\left( 2\,M - r \right) \,\cos (\theta )\,
       {F_{12}}(t,r,\theta ,\phi )
\nonumber \\ &&+
r\,\left( \cos (\theta )\,{F_{21}}(t,r,\theta ,\phi ) +
         {\sqrt{1 - \frac{2\,M}{r}}}\,
          \left( -\left( \sin (\theta )\,
               {F_{31}}(t,r,\theta ,\phi ) \right)  +
            8\,{{\omega }_4}(t,r,\theta ,\phi ) \right)  \right),
\nonumber \\ &&
G_{ab}[3, 4, 1,2,3]=\frac{-{F_{41}}(t,r,\theta ,\phi ) -
      8\,{{\omega }_3}(t,r,\theta ,\phi )}{2},
\nonumber \\ &&
G_{ab}[4, 1, 2,3,4]=\frac{r\,\cos (\theta )\,
       {F_{12}}(t,r,\theta ,\phi ) -
      \sin (\theta )\,{F_{13}}(t,r,\theta ,\phi ) -
      \frac{8\,{\sqrt{1 - \frac{2\,M}{r}}}\,r^2\,
         {{\omega }_4}(t,r,\theta ,\phi )}{-2\,M + r}}{2} ,
\nonumber \\ &&
G_{ab}[4, 1, 1,3,4]=\frac{-\left( r\,\cos (\theta )\,
{F_{11}}(t,r,\theta ,\phi ) \right) }{2} ,
\nonumber \\ &&
G_{ab}[4, 1, 1,2,4]=\frac{\sin (\theta )\,
       {F_{11}}(t,r,\theta ,\phi ) +
      \frac{M\,{F_{44}}(t,r,\theta ,\phi )}
       {{\sqrt{1 - \frac{2\,M}{r}}}\,r^2}}{2} ,
\nonumber \\ &&
G_{ab}[4, 1, 1,2,3]=\frac{-\left( M\,
         {F_{43}}(t,r,\theta ,\phi ) \right)  +
      8\,r^3\,{{\omega }_1}(t,r,\theta ,\phi )}{2\,
      {\sqrt{1 - \frac{2\,M}{r}}}\,r^2} ,
\nonumber \\ &&
G_{ab}[4, 2, 2,3,4]=\frac{r\,
      \left( \cos (\theta )\,{F_{22}}(t,r,\theta ,\phi ) -
        {\sqrt{1 - \frac{2\,M}{r}}}\,\sin (\theta )\,
         {F_{32}}(t,r,\theta ,\phi ) \right) }{2},
\nonumber \\ &&
G_{ab}[4, 2, 1,3,4]=\frac{1}{2\,r}\left( 2\,M - r \right) \,\sin (\theta )\,
{F_{13}}(t,r,\theta ,\phi )
\nonumber \\ &&+
      r^2\,\left( -\left( \cos (\theta )\,
            {F_{21}}(t,r,\theta ,\phi ) \right)  +
         {\sqrt{1 - \frac{2\,M}{r}}}\,
          \left( \sin (\theta )\,{F_{31}}(t,r,\theta ,\phi ) -
            8\,{{\omega }_4}(t,r,\theta ,\phi ) \right)  \right),
\nonumber \\ &&
G_{ab}[4, 2, 1,2,4]=
\frac{\left( 1 - \frac{2\,M}{r} \right) \,\sin (\theta )\,
      {F_{12}}(t,r,\theta ,\phi )}{2},\nonumber \\ &&
G_{ab}[4, 2, 1,2,3]=4\,{\sqrt{1 - \frac{2\,M}{r}}}\,r\,
    {{\omega }_2}(t,r,\theta ,\phi ) ,
\nonumber \\ &&
G_{ab}[4, 3, 2,3,4]=\frac{\frac{\cos (\theta )\,
         {F_{23}}(t,r,\theta ,\phi )}{{\sqrt{1 - \frac{2\,M}{r}}}}
       - \sin (\theta )\,{F_{33}}(t,r,\theta ,\phi ) -
      {F_{44}}(t,r,\theta ,\phi )}{2},
\nonumber \\ &&
G_{ab}[4, 3, 1,3,4]=\frac{-\left( {\sqrt{1 - \frac{2\,M}{r}}}\,
        \cos (\theta )\,{F_{13}}(t,r,\theta ,\phi ) \right) }{2},
\nonumber \\ &&
G_{ab}[4, 3, 1,2,4]=\frac{1}{2\,{\sqrt{1
- \frac{2\,M}{r}}}\,r}\left( -2\,M + r \right) \,
       \cos (\theta )\,{F_{12}}(t,r,\theta ,\phi )
\nonumber \\
&&+ r\,\left( -\left( \cos (\theta )\,
{F_{21}}(t,r,\theta ,\phi ) \right)  +
{\sqrt{1 - \frac{2\,M}{r}}}\,
\left( \sin (\theta )\,{F_{31}}(t,r,\theta ,\phi ) -
8\,{{\omega }_4}(t,r,\theta ,\phi ) \right)  \right),
\nonumber \\ &&
G_{ab}[4, 3, 1,2,3]=\frac{{F_{41}}(t,r,\theta ,\phi ) +
8\,{{\omega }_3}(t,r,\theta ,\phi )}{2}.
\label{so}
\end{eqnarray}
We note now that $F_{21}=F^2_1,F_{31}=F^3_1$
can be the only non-vanishing components compatible with (\ref{so}),
i.e. $\omega_\mu=0$ and
$F_{ij}=0$ with $i\neq 2,3$ and $j\neq 1$. To satisfy the field
equations one should have $F_{21}=F_{31}=0$, because (cf. Sec. VI)
\begin{eqnarray}
&&K_{1, 1,3,4}=\frac{-8\,M\,\sin (\theta )\,
{F_{21}}(t,r,\theta ,\phi )}{r},
\nonumber \\ &&
K_{1, 1,2,4}=\frac{4\,M\,\sin (\theta )\,
{F_{31}}(t,r,\theta ,\phi )}{{\sqrt{1 - \frac{2\,M}{r}}}\,r^2},
\end{eqnarray}
but if we do so we fail to satisfy that part of the torsion
constraint that involves the component along the Minkowski metric
$\eta_{ab}$, i.e. ${\widetilde T}^{a}\wedge V^{b}$, which requires
having at least $F_{31}=F^3_1\neq 0$. For example, one of the
components is given by
\begin{eqnarray}
\; & \; &
8\,\biggr( -\left( {\Lambda }^{23}\,\cos (\theta ) \right)  +
\frac{6\,M\,{\Lambda }^{23}\,\cos (\theta )}{r} +
r\,({{F_1}}^3)^{(0,0,0,1)}(t,r,\theta ,\phi )
\nonumber \\
&-&{\sqrt{1 - \frac{2\,M}{r}}}\,
({F_3})^{(0,0,0,1)}(t,r,\theta ,\phi)
-r\,\sin (\theta )\,({{F_1}}^4)^{(0,0,1,0)}(t,r,\theta ,\phi)
\nonumber \\
&+& {\sqrt{1 - \frac{2\,M}{r}}}\,
({F_4})^{(0,0,1,0)}(t,r,\theta ,\phi ) +
r\,\sin (\theta )\,({{F_3}}^4)^{(1,0,0,0)}(t,r,\theta,\phi)
\nonumber \\
&-& r\,({{F_4}}^3)^{(1,0,0,0)}(t,r,\theta,\phi)\biggr)=0,
\end{eqnarray}
where the four upstairs indices separated by commas denote how many
derivatives are taken with respect to $t,r,\theta,\phi$, respectively.

\acknowledgments

E.  Di Grezia and G. Esposito are grateful to the Dipartimento di Scienze Fisiche
of Federico II University, Naples, for hospitality and support. We are
indebted to Paolo Aschieri for enlightening correspondence.

\end{document}